%% file: manuscript.tex
\documentclass[twocolumn, emulateapj]{aastex63}
\usepackage[utf8]{inputenc}
\usepackage{gensymb}

\usepackage{natbib}
\usepackage{float}
\usepackage{ulem}
\usepackage{relsize}
\pdfoutput=1
\usepackage{color}
\usepackage{amsmath}
\usepackage{amsthm}
\usepackage{microtype}

\newcommand{\Mj}{$M_{\mathrm{J}}$}

\newenvironment{packed_enum}{
\begin{enumerate}
  \setlength{\itemsep}{1pt}
  \setlength{\parskip}{0pt}
  \setlength{\parsep}{0pt}
}{\end{enumerate}}

\begin{document}

\title{Radio Emission from Binary Ultracool Dwarf Systems}

\author[0000-0001-5125-1414]{Melodie M. Kao}
\affil{University of California Santa Cruz, Department of Astronomy \& Astrophysics, 1156 High Street, Santa Cruz, CA 95064 USA}
\affil{Arizona State University, School of Earth and Space Exploration, 781 Terrace Mall, Tempe, AZ 85287 USA}
\affil{Heising-Simons Foundation 51 Pegasi b Fellow}
\affil{NASA Hubble Postdoctoral Fellow}
\email{melodie.kao@ucsc.edu}
\author[0000-0002-4489-0135]{J.~Sebastian Pineda}
\affiliation{University of Colorado Boulder, Laboratory for Atmospheric and Space Physics, 3665 Discovery Drive, Boulder CO, 80303, USA}


\begin{abstract}
\input{SECTION_abstract}

\end{abstract}




\section{Introduction}\label{sec.intro}

\input{SECTION_intro}

\input{SECTION_newObservations}

\input{SECTION_discussion}

\input{SECTION_conclusions}

\acknowledgments
MK especially thanks Jackie Villadsen for her help in troubleshooting target calibration and imaging. Support was provided by NASA through the NASA Hubble Fellowship grant HST-HF2-51411.001-A awarded by the Space Telescope Science Institute, which is operated by the Association of Universities for Research in Astronomy, Inc., for NASA, under contract NAS5-26555; and by the National Radio Astronomy Observatory.  The National Radio Astronomy Observatory is a facility of the National Science Foundation (NSF) operated under cooperative agreement by Associated Universities, Inc. This work is based on observations made with the NSF's Karl G. Jansky Very Large Array (VLA).  This research has made use of the SIMBAD and VizieR databases, operated at CDS, Strasbourg, France; and the European Space Agency (ESA) mission \textit{Gaia} (\url{https://www. cosmos.esa.int/gaia}), processed by the Gaia Data Processing and Analysis Consortium (DPAC, \url{https://www. cosmos.esa.int/web/gaia/dpac/consortium}).

\facility{JVLA}
\software{CASA \, \citep{CASA}, Astropy \, \citep{astropy:2018},\, Matplotlib \citep{matplotlib},\, Numpy \,\citep{numpy2}, \, Scipy \, \citep{scipy}}

\clearpage
\bibliography{occurrenceRate}


\end{document}

%% file: SECTION_abstract.tex
Well-characterized binary systems will provide valuable opportunities to study the conditions that are necessary for the onset of both auroral and non-auroral magnetospheric radio emission in the ultracool dwarf regime. We present new detections of non-auroral ``quiescent" radio emission at 4--8 GHz of the three ultracool dwarf binary systems GJ~564~BC, LP~415-20, and 2MASS~J21402931+1625183. We also tentatively detect a highly circularly polarized pulse at 4--6 GHz that may indicate aurorae from GJ~564~BC. Finally, we show that the brightest binary ultracool dwarf systems may be more luminous than predictions from single-object systems.

%% file: SECTION_intro.tex
Since the first discovery of radio emission \citep{Berger2001Natur.410..338B} from ultracool dwarfs with M7 and later spectral types, GHz radio observations of such objects have revolutionized our understanding of how magnetic activity evolves over the star-planet transition regime. Surveys of X-ray emission in the lowest-mass stars showed a sharp drop-off in coronal emissions at a spectral type M9.5 \citep{Williams2014ApJ...785....9W}, with chromospheric H$\alpha$ emission declining continuously across the L-dwarf regime \citep{Schmidt2015AJ....149..158S, Pineda2016ApJ...826...73P}. Despite these declining diagnostics of magnetic activity, radio observations demonstrated persistently strong GHz emissions across the entire ultracool dwarf regime, revealing the emergence of radio aurorae \citep{Hallinan2015Natur.523..568H, Kao2016ApJ...818...24K, Pineda2017ApJ...846...75P}.

The radio component of ultracool dwarf aurorae manifests as periodically flaring and highly circularly polarized coherent electron cyclotron maser emission \citep{Hallinan2007ApJ...663L..25H, Hallinan2008ApJ...684..644H}.  This emission traces the fundamental cyclotron frequency of the local magnetosphere \citep{Treumann2006AARv..13..229T}. Detections of GHz radio aurorae confirm that ultracool dwarfs  at least as late as T6.5 can host strong kiloGauss magnetic fields \citep{RouteWolszczan2012ApJ...747L..22R, RouteWolszczan2016ApJ...821L..21R, Williams2015ApJ...808..189W, Kao2016ApJ...818...24K, Kao2018ApJS..237...25K, Richey-Yowell2020}.

Ultracool dwarfs also exhibit nonthermal and incoherent radio emission that is quasi-steady and weakly circularly polarized \citep[e.g.,][]{Williams2015ApJ...799..192W, Kao2016ApJ...818...24K, Kao2018ApJS..237...25K}.  This  ``quiescent" radio emission has been detected at frequencies as high as 95 GHz \citep{Williams2015ApJ...815...64W, hughes2021AJ....162...43H} and is attributed to optically thin gyrosynchrotron emission \citep[e.g.][]{Berger2005ApJ...627..960B, Osten2006ApJ...637..518O, Williams2015ApJ...815...64W, Lynch2016MNRAS.457.1224L}. Additionally, it can persist for years \citep[e.g.][]{Berger2008ApJ...676.1307B, Kao2016ApJ...818...24K, Kao2018ApJS..237...25K}. Furthermore, quiescent radio luminosities correlate with H$\alpha$ luminosities in aurorae-emitting ultracool dwarfs, suggesting that the physical conditions underpinning ultracool dwarf radio aurorae may also be related to their quiescent radio emission \citep{Pineda2017ApJ...846...75P}.  \citet{Kao2019MNRAS.487.1994K} argue that one plausible explanation is radiation belts analogous to the extended circum-planetary regions of high-energy electrons trapped in the magnetospheres of solar system planets \citep{sault1997, bolton2004, Clarke2004jpsm.book..639C, Horne2008}. While the source of this magnetospheric plasma is unknown,  flares  \citep[e.g.][]{Gizis2013ApJ...779..172G, Paudel2018ApJ...858...55P, Jackman2019MNRAS.485L.136J, Paudel2020arXiv200410579P}  are one possibility, and \citet{Kao2018ApJS..237...25K, Kao2019MNRAS.487.1994K} speculate that volcanic activity from planets could be another. 

In this work, we examine radio emission in three ultracool dwarf binary systems.  
Such systems will provide valuable tests of ultracool dwarf magnetic activity since unlike the field population, binaries can have precise observational constraints on individual masses and therefore ages \citep[e.g.][]{Konopacky2010ApJ...711.1087K, Dupuy2017ApJS..231...15D, Dupuy2019AJ....158..174D}.  Precisely known properties enable robust comparisons between individual objects and the conditions which power their radio emission.  For instance, in two radio ultracool dwarf binary systems, only one component exhibits detectable GHz radio emission \citep{Konopacky2012ApJ...750...79K,Harding2013A&A...554A.113H,Williams2015ApJ...799..192W, Forbrich2016ApJ...827...22F}.  
Why is one component radio bright but the other is not?   Additional observations of well-characterized systems, such as those presented in this work, will help elucidate this question.

%% file: SECTION_newObservations.tex
\input{TABLE_targets}

\section{Targets}\label{sec:targets}
We present new observations of three binary systems using the \textit{NSF's Karl G. Jansky} Very Large Array \citep[VLA;][]{Perley2011ApJ...739L...1P} from program VLA 18B-283 (PI - Pineda). We selected these binary systems for their fast rotation. \citet{Pineda2017ApJ...846...75P} showed that the fraction of ultracool dwarfs detected at radio frequencies rise as a function of projected rotational velocity $v\sin i$ for speeds $\gtrsim$35 km s$^{-1}$, which corresponds to a $\sim$3.5-hour period when viewed at an inclination of 90$^{\degree}$. This strong dependence on rotation is consistent with existing theories for driving the electrodynamic engines of aurorae \citep[e.g.][]{Nichols2012ApJ...760...59N, Turnpenney2017MNRAS.470.4274T} and may be a requirement for generating the strong dipolar magnetic field topologies \citep{Shulyak2017NatAs...1E.184S} that are a critical ingredient to powering ultracool dwarf auroral processes \citep{Pineda2017ApJ...846...75P}.

Table \ref{table:targets} summarizes the properties of our targets and we discuss them in further detail below.

\textbf{GJ 564 BC} is also known as HD~130948BC, a benchmark brown dwarf system orbiting a solar analog \citep{Dupuy2009ApJ...692..729D}. It was discovered through adaptive optics imaging \citep[][]{Potter2002ApJ...567L.133P} and is the nearest binary in our observation sample.  This $0.44\pm0.04$ Gyr system consists of two nearly equal-mass L4 dwarfs on a $\sim$10-year orbit with masses of $59.8^{+2.0}_{-2.1}$ and $55.6^{+2.0}_{-1.9}$ \Mj\ for the B and C components, respectively \citep{Dupuy2017ApJS..231...15D}.  This system is unresolved from GJ~564 A except in adaptive optics imaging. H$\alpha$ emission for this system is unknown because the binary is unresolved in spectroscopic observations from the solar type primary.

\textbf{2MASS J21402931+1625183} was first identified as a binary system by \citet{Close2002ApJ...567L..53C,Close2003ApJ...587..407C}. Earlier attempts to determine masses for this system suggested a large primary to secondary mass ratio \citep[$\sim$4;][]{Konopacky2010ApJ...711.1087K}. Most recently, \citet{Dupuy2017ApJS..231...15D} reported that the primary has a mass of $114_{-12}^{+10}$ \Mj\, and the secondary has a mass of  $69_{-9}^{+8}$ \Mj. These updated values confirm the suggestion by \citet{Konopacky2010ApJ...711.1087K} that their estimated mass ratio was likely too high.  \citet{Dupuy2017ApJS..231...15D} did not provide an age estimate for this system, but \citet{Martin2017ApJ...838...73M} reported that gravity-sensitive indices in the near-IR spectrum of this system indicate that it is not young (FLD-G classification, $\gtrsim$200 Myr). 
\citet{Gizis2000AJ....120.1085G} report an H$\alpha$ EW of 0 \AA.  This low-eccentricity system ($e= 0.196\pm0.007$) has a $\sim$24.4-year orbit. 

\textbf{LP 415-20} was first  identified as a binary by \citet{Siegler2003ApJ...598.1265S}. It is another unequal-mass binary system with masses of $156_{-18}^{+17}$ \Mj\, and $92_{-18}^{+16}$ \Mj\, for the primary and secondary, respectively \citep{Dupuy2017ApJS..231...15D}.  However, \citet{Dupuy2017ApJS..231...15D} note that the mass of the primary component is much higher than expected for its luminosity and speculate that the primary may be an unresolved unequal-mass binary. If so, the primary would likely be comprised of clearly ultracool dwarf $\sim$100 \Mj\ and $\sim$50 \Mj\ components, whose combined light are estimated as an M6 object, although spectral decomposition suggests spectra are consistent with templates of types M5-M7. This high eccentricity ($e = 0.706_{-0.012}^{+0.011}$) system has a $\sim$14.8-year orbit. The model-derived age of the secondary component is $5.0_{-4.7}^{+1.9}$ Gyr, and an unresolved binary system for the primary component points to a model-derived age of at least several hundred Myr. Using BANYAN $\Sigma$ \citep{Gagne2018ApJ...856...23G}, we find that LP~415-20 is a likely member of the Hyades Cluster with $\geq$99\% probability  \citep[$750 \pm 100$;][]{Brandt2015ApJ...807...58B} when using the \textit{Gaia} parallax and proper motion data along with a mean system radial velocity of $40.8 \pm 1.4$ km s$^{-1}$ \citep{Konopacky2010ApJ...711.1087K}. 
 \citet{MilesPaez2017MNRAS.472.2297M} report significant photometric variability in the combined light of the system with a periodicity of $\sim$4.36 hrs. The system also shows H$\alpha$ in emission, with an EW measurement of 4.4 \AA\ \citep{Gizis2000AJ....120.1085G}.

\section{Observations and Calibrations}
\input{TABLE_observations}

We summarize target observations in Table \ref{table:obs}. For GJ~564~BC, we obtained two 2-hr observing blocks with the VLA for four total hours on sky.  For 2MASS~J21402931+1625183 and LP 415-20, we obtained one 5-hr observing block each.

\begin{figure*}[!ht]
\epsscale{1.2}
\plotone{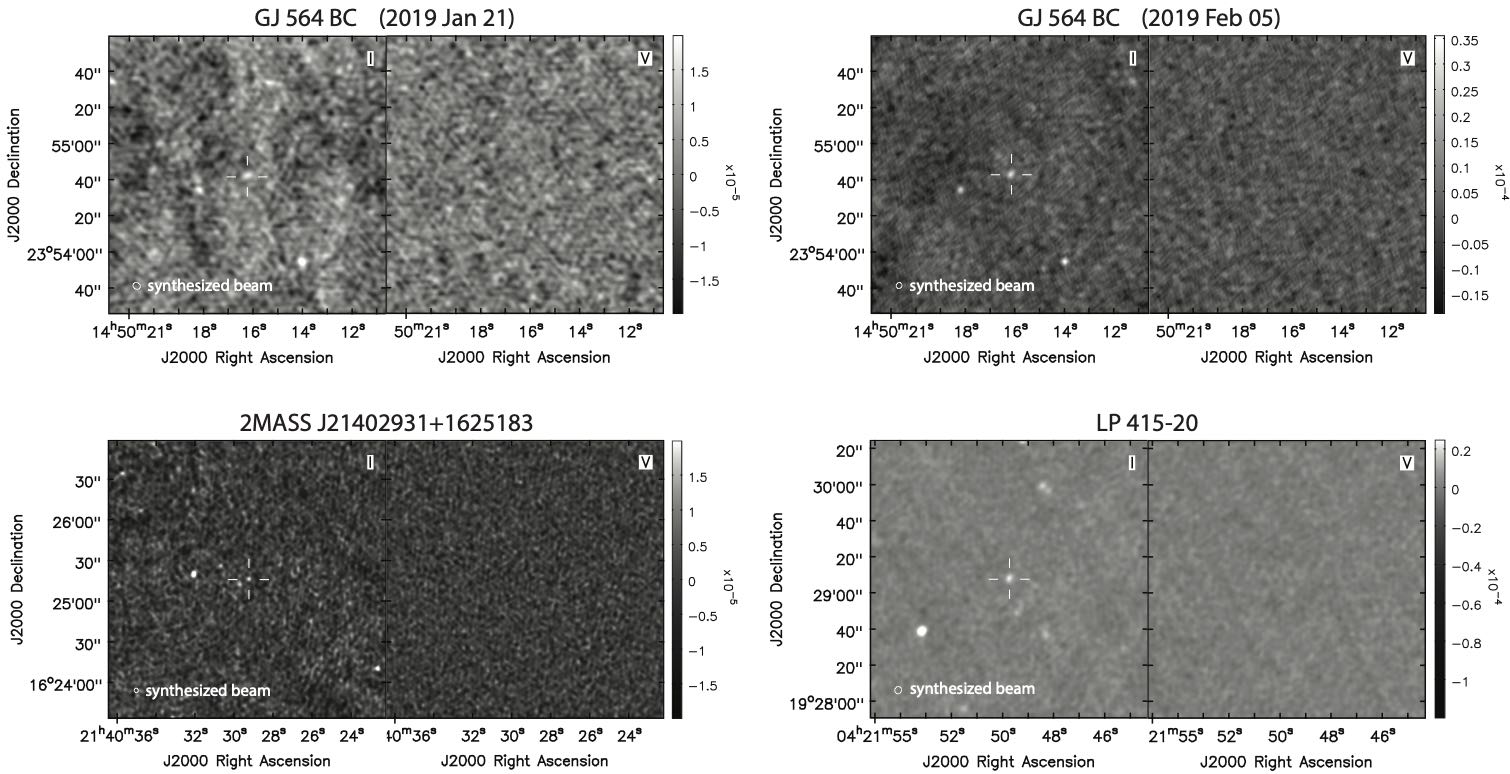}
\caption{\label{fig:imaging}  4--8 GHz Stokes I and V images averaged over the full observing block for each target.  Cross hairs indicate detected flux at the expected locations of our targets.  No Stokes V emission was detected from our targets.}
\end{figure*}

We calibrated our measurement sets using nearby phase calibrators and the standard VLA flux calibrators 3C147 and 3C286.  Typical full-bandwidth sensitivities at C configuration for 2- and 5-hr integration blocks with 3.5 GHz bandwidth (to account for RFI excision) centered at 6.0 GHz are 2.1\,$\mu$Jy and  1.4\,$\mu$Jy, respectively, and reach absolute flux calibration accuracy of $\sim$5\%.

To account for phase errors that can systematically reduce flux densities, the National Radio Astronomy Observatory (NRAO) recommended phase calibration cycle times of $\sim$25 minutes for observations at 4--8 GHz in C configuration when these observations took place.  We adhered to these guidelines for our observations.  For GJ~564~BC, we alternated between a nearby phase calibrator and the target with  integration times of 2 and 23.75 minutes, respectively, for a total cycle time of 25.75 minutes.  For 2MASS J21402931+1625183, these integration times were 2 and 20.6 minutes, for a total cycle time of 22.6 minutes.  For LP~415-20, the integration times were 2 and 23.5 minutes, for a total cycle time of 25.5 minutes.

Sidelobes from a bright $\sim$31.3 mJy object located $\sim$3.3 arcmin to the northeast  lead to poor initial image quality for 2MASS J21402931+1625183. We self-calibrate its target field using this bright object to improve our image rms noise by a factor of 12, from 43.4 $\mu$Jy to  3.6 $\mu$Jy.

We did not observe polarization calibrators, but the absence of polarization calibration is not a limiting factor for our analysis.  Polarization leakage at typical levels of 2--3\% result in spurious Stokes V (circularly polarized) flux densities of $\sim$1.2 $\mu$Jy for our brightest source, which is less than the noise floor.

We initially processed each measurement set with the VLA CASA 5.6.2 Calibration Pipeline, after which we flagged all remaining radio frequency interference (RFI) and checked all calibrations.  As a rule, all data between 4.0--4.4~GHz was discarded due to extremely bright and persistent RFI. We obtained absolute flux by bootstrapping flux densities with the observed flux calibrators.

\input{TABLE_imaging}

\section{Imaging}

We produced Stokes I and Stokes V (total and circularly polarized intensities, respectively) images for the entire observing block of each object  with the CASA \texttt{tclean} routine.  For GJ~564~BC and LP 415-20, we model the frequency dependence of sources with three terms to account for curvature in the spectrum.  We also use natural weighting for best point-source sensitivity, multi-scale cleaning with a bias of 0.5 to more heavily weight point sources, and set the cell size to $0\farcs3 \times 0\farcs3$.  For 2MASS J21402931+1625183, dynamic range limits its imaging.  To suppress sidelobes from bright sources in the field of view, we use Briggs weighting and set \texttt{robust = 0.5}.  Bright sources outside of the primary beam cause image artifacts due to non-coplanar baselines. We mitigate these artifacts with w-projection and reduce some computational overhead by slightly increasing the cell size to $0\farcs4 \times 0\farcs4$. Finally, we model the frequency dependence of sources with four terms to account for artifacts that cannot be satisfactorily modeled with three terms.

We searched for a point source at the proper motion-corrected coordinates of each target by eye and detect radio emission from all three of our targets, including for both epochs of GJ~564~BC. 

For each object, we fit the flux density of the source using two methods.  Prior to flux fitting, we added phase delays to our visibility data to transform the phase center of our data to the measured locations of our targets using the CASA task \texttt{fixvis}. Shifting to the phase center allows for flux density measurements with the CASA task \texttt{uvmodelfit}, which requires that the target be at or very near the phase center.

 After cleaning the target field, we subtracted models for all sources except for the target using the  CASA task \texttt{uvsub}.  Then, we fit the calibrated visibilities with a point source model using the CASA task \texttt{uvmodelfit}.  We also fit an elliptical Gaussian point source to the cleaned image of each object using the CASA task \texttt{imfit}. Table \ref{table:imaging} gives measured flux densities and percent circular polarizations for detected objects and 3$\sigma$ upper limits on the flux densities for undetected objects. Figure \ref{fig:imaging} shows Stokes I and V images for all targets.

\section{Timeseries}

\begin{figure*}[!t]
\epsscale{1.1}
\plotone{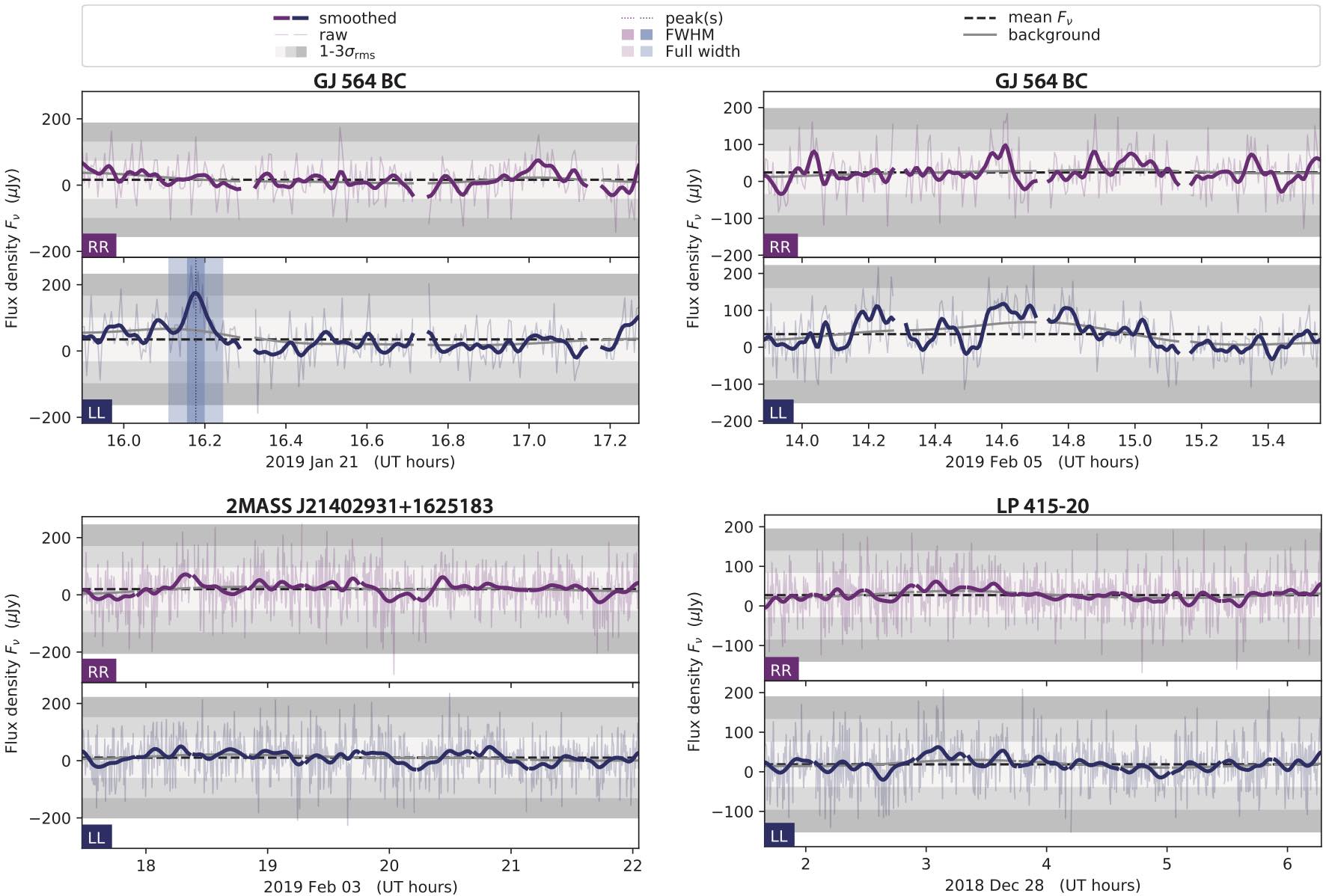}
\caption{\label{fig:timeseries}  4--8 GHz right- and left-circularly polarized (RR and LL, respectively) timeseries for detected objects with 10s averaging.  Gaps in timeseries correspond to phase calibration scans and rms values shown in grey correspond to the rms of the raw time-averaged data. A left circularly polarized candidate pulse was detected in the 2019 Jan 21 epoch for GJ 564 BC, and shaded regions around that pulse show the calculated FWHM and full width regions.    }
\end{figure*}

We also generated timeseries for each object's right- and left-circularly polarized  (rr- and ll-correlations) emission.  To do this, we time-averaged and frequency-averaged the phase-centered data into 10s, 30s, and 60s time resolutions each for 4--8~GHz, 4--6~GHz, and 6--8~GHz sub-bands using the averaging functions available with CASA's plotting tool \texttt{plotms}, and we exported the averaged timeseries of the real component of the phase-centered visibilities. Using these different time resolutions, we check for possible substructure in identified candidate pulses (see below) and also check that such candidate pulses are present for all time resolutions. 

We then searched these timeseries for candidates of highly circularly polarized pulses indicative of auroral emission using the following revision of the procedure described in \citet{Kao2018ApJS..237...25K}:
\begin{packed_enum}
    \item Calculate a timeseries $t_q$ for slowly varying quiescent emission by smoothing the raw timeseries with a Gaussian kernel that has a width of  10\% of the total time on-source. 
    \item Construct a timeseries $t_p$  for pulse-searching by smoothing the raw timeseries with a Gaussian kernel that has a width of  1\% of the total time on-source.  This mitigates residual noise spikes that may not have been identified for excision by both the CASA data reduction pipeline and our own manual examination of the data, which are much narrower than observed radio aurorae pulses that have durations at least as long as several minutes \citep[e.g.][]{Berger2001Natur.410..338B, BurgasserPutnam2005ApJ...626..486B, Hallinan2007ApJ...663L..25H, Hallinan2008ApJ...684..644H, Berger2006ApJ...648..629B, Berger2009ApJ...695..310B, RouteWolszczan2012ApJ...747L..22R, RouteWolszczan2016ApJ...830...85R, Williams2015ApJ...799..192W,  Gizis2016AJ....152..123G, Kao2016ApJ...818...24K, Kao2018ApJS..237...25K, Zhang2020ApJ...897...11Z}.  
    \item Subtract $t_q$ from $t_p$ to obtain a residual timeseries $t_r$  without the quiescent component. 
    \item Identify peaks with at least 2$\sigma_{\text{rms}}$, where $\sigma_{\text{rms}}$ is the root mean square of $t_r$. This lower significance accounts for the possibility that $t_r$ may have pulses that could elevate $\sigma_{\text{rms}}$ and prevent the identification of weaker pulses.  Thus, we remove the strongest peaks in the first iteration of this procedure and then repeat it with an updated  $\sigma_{\text{rms}}$ that excludes any initially identified peaks.
    \item Calculate the full-width half-maximum (FWHM) of identified peaks. 
    \item Remove the full width of each identified peak from $t_r$, where we define the full width of each peak as three times the FWHM.  This gives the updated residual timeseries $t_{r,u}$. 
    \item Repeat steps 4--6, using  $t_{r,u}$ for calculating $\sigma_{\text{rms}}$ and $t_{r}$ for peak-searching.  For this iteration, we also require a peak significance threshold of 3$\sigma_{\text{rms}}$.  Returned peaks are candidate pulses.  Note that by using $t_{r}$ for the peak search, this procedure does not inherit identified peaks using the lower significance from the first iteration.  Instead, all candidate pulses identified by the full procedure must meet a 3$\sigma_{\text{rms}}$ significance.
\end{packed_enum}
We emphasize that the sole function of the procedure outlined above is to aid in identifying candidate pulses in a repeatable manner. Final confirmation of identified candidate pulses relies on human judgement. Figure \ref{fig:timeseries} shows the 4--8~GHz timeseries for each object. 

\begin{figure}[!t]
\epsscale{1.1}
\plotone{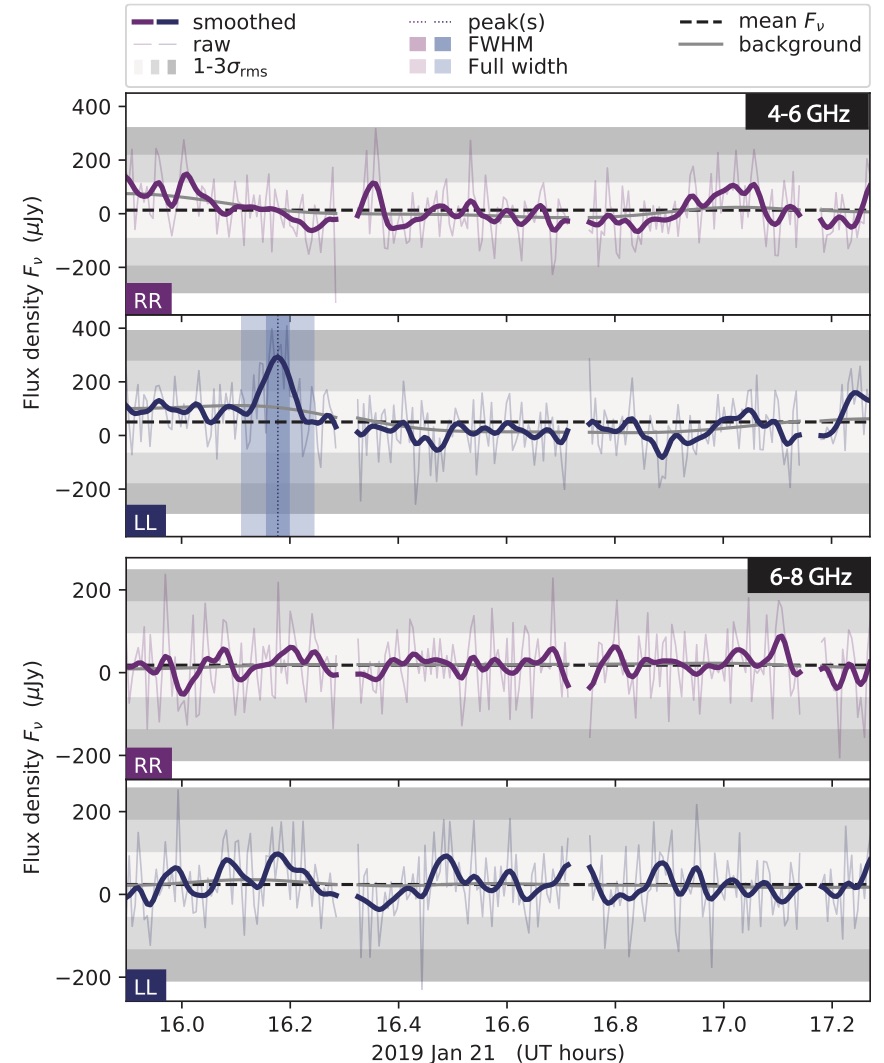}
\caption{\label{fig:timeseries_gj564}  4--6 GHz and 6--8 GHz right- and left-circularly polarized (RR and LL, respectively) timeseries for GJ 564 BC. Shaded regions around that pulse show the calculated FWHM and full width regions. The candidate pulse appears to drop out at 6--8 GHz. }
\end{figure}

We detect one candidate pulse from GJ 564 BC during its first epoch which is present for all timeseries resolutions. Comparing the 4--6 GHz and 6--8 GHz timeseries shows that the pulse appears only in the lower subband (Figure \ref{fig:timeseries_gj564}).
To confirm this pulse, we image over the time interval of its FWHM at 4--6 GHz and measure its Stokes I ($95\pm27$ $\mu$Jy ) and Stokes V ($125\pm24$ $\mu$Jy) flux densities. While Stokes V flux densities cannot physically exceed Stokes I flux densities, the measured Stokes I and V flux densities are consistent within $3\sigma_{\mathrm{rms}}$ and correspond to a lower bound percent circular polarization of $\geq$38\% with 99.7\% confidence.
For comparison, imaging outside of the full width of the identified pulse using the full bandwidth yields a Stokes I quiescent emission flux density of $19.2 \pm 4.9$ $\mu$Jy, and no Stokes V source is detected to a $3\sigma_{\mathrm{rms}}$ significance of $\leq$15.0 $\mu$Jy.   

\begin{figure}[!t]
\epsscale{1.1}
\plotone{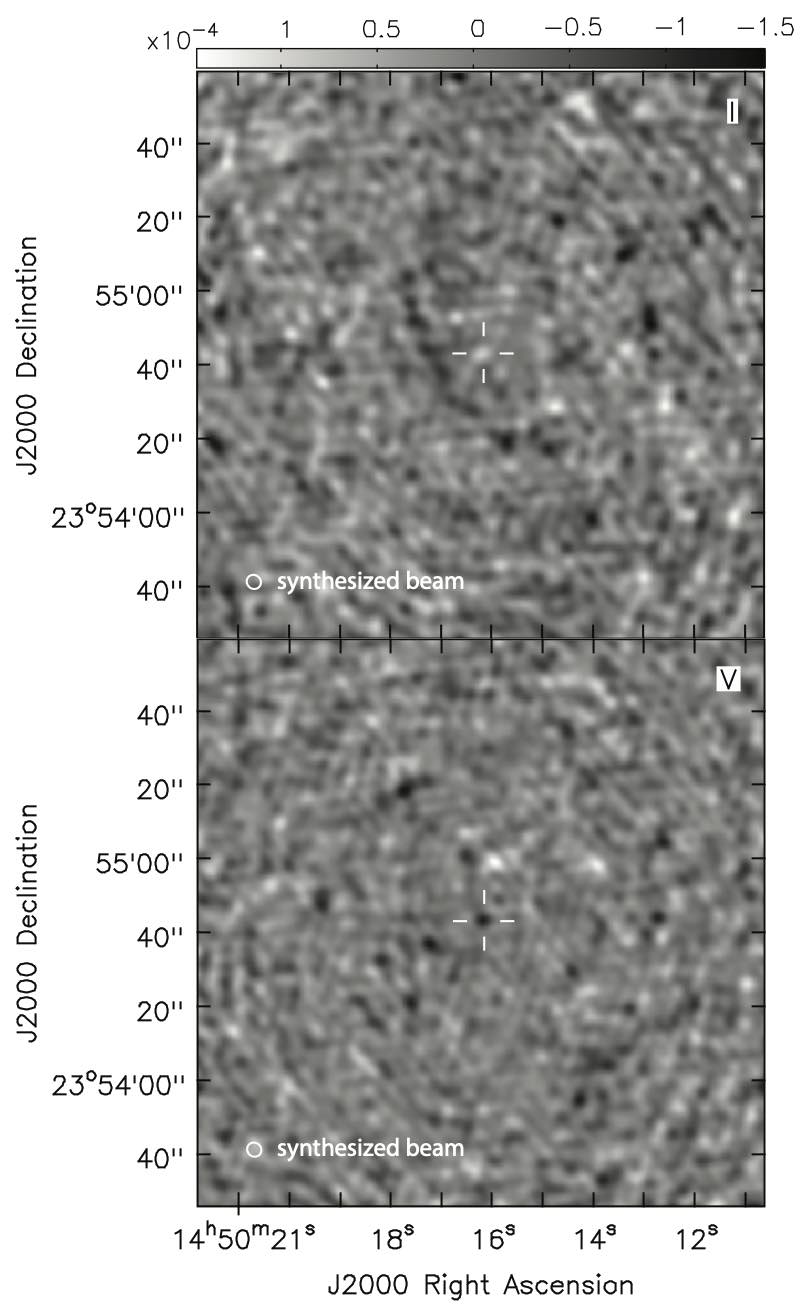}
\caption{\label{fig:imaging_gj564_pulse} Image of GJ 564 BC over the FWHM of its identified candidate pulse.  The target is visually indistinguishable from noise peaks in Stokes I and marginally distinguishable in Stokes V. Bootstrapping yields a significance of 98.5\% for the  Stokes I flux density and 99.9\% for the Stokes V flux density. We classify this pulse as tentative rather than confirmed. }
\end{figure}

However, the target is difficult to distinguish from noise peaks in the image (Figure \ref{fig:imaging_gj564_pulse}), so we bootstrap the significance of the measured Stokes I and Stokes V flux densities by fitting a point source in 10,000 randomly drawn fitting regions of size 50$\times$50 pixels.  We find that the Stokes I flux density has a significance of 98.5\% and the Stokes V flux density has a significance of 99.9\%.  We therefore classify this candidate pulse as only a tentative detection.   Similarly, the Stokes I flux density of the quiescent emission is also tentative, corresponding to a 99.8\% significance. 

We did not identify any candidate pulses from 2MASS J21402931+1625183 and LP 415-20 despite covering their likely full rotation periods.  However, we note that our chosen time-averaging is only sensitive to pulses that are at least as bright as a peak flux of $\sim$100--200 $\mu$Jy, or $\sim$5--10$\times$ brighter than their measured quiescent emission.  Fainter pulses may exist to which our observations are not sensitive.

%% file: TABLE_targets.tex
\setlength{\tabcolsep}{0.045in}
\begin{deluxetable*}{llccccccl}
\tabletypesize{\scriptsize}
\tablecaption{Targets \label{table:targets}}
\tablehead{
\colhead{ Object }	&	\colhead{SpT\tablenotemark{\tiny{c}}} 	&	\colhead{$v \sin i$\tablenotemark{\tiny{c}}} 	& \colhead{$a$} 	&	\colhead{ $\pi$ }	&	\colhead{ d }		&	\colhead{ $\mu_{\alpha} \cos \delta$ }	&	\colhead{ $\mu_{\delta}$ }	&	\colhead{ ref }	\\[-9pt]
\colhead{}				&	\colhead{}		&	\colhead{(km s$^{-1}$)} & \colhead{(AU)}	&	\colhead{ (mas) }	&	\colhead{ (pc) }	&	\colhead{ (mas yr$^{-1}$) }				&	\colhead{ (mas yr$^{-1}$) }	&	\colhead{}
}
\startdata
GJ 564 BC\tablenotemark{\tiny{a}}   &	L4.0 $\pm$ 1.0 ; L4.0 $\pm$ 1.0	& 62 $\pm$ 4 ; 86 $\pm$ 6	&	2.226 $^{+0.014}_{-0.013}$	&	54.9068 $\pm$ 0.0684	&	18.2127	$\pm$	0.0227	&	\phm{-}144.7\phn\phn $\pm$ 0.8\phn\phn	&	\phm{-}32.4\phn\phn	$\pm$ 0.7\phn\phn	&	1 4 3 2	\\
J2140+16\tablenotemark{\tiny{b}}	&	M8.0 $\pm$ 0.5 ; L0.5 $\pm$ 1.0	& 13 $\pm$ 2 ; 37 $\pm$ 3 	&	4.71 $\pm$ 0.14	            &	30.1972	$\pm$ 0.434\phn	&	33.1	$\pm$	0.5	    &	\phn-77.9\phn\phn $\pm$	0.776			&	-85.637	$\pm$ 0.707	&	1 4	3	\\
LP 415$-$20	                        &	M6.0 $\pm$ 1.0 ; M8.0 $\pm$ 0.5	& 40 $\pm$ 5 ; 37 $\pm$ 4	&	3.73 $\pm$ 0.12	        	&	25.1963	$\pm$ 0.5117	&	39.6884	$\pm$	0.8060	&	\phm{-}134.716 $\pm$ 0.912				&	-38.416	$\pm$ 0.619	&	1 4	3
\enddata
\tablenotetext{a} {Also known as HD 130948B}
\tablenotetext{b} {2MASS J21402931+1625183}
\tablenotetext{c} {Listed values are for the primary and secondary, respectively.}
\tablerefs{
(1)				\citep{Dupuy2017ApJS..231...15D}	;
(2)				\citep{Faherty2009AJ....137....1F}	;
(3)				\citep{Gaia2018yCat.1345....0G}		;
(4)				\citep{Konopacky2012ApJ...750...79K}}
\end{deluxetable*}

%% file: TABLE_observations.tex
\setlength{\tabcolsep}{0.03in}
\begin{deluxetable}{lcccccccc}[!ht]
\tabletypesize{\scriptsize}
\tablecaption{Summary of observations \label{table:obs}}
\tablehead{	
\colhead{}	            & \colhead{}	       & \colhead{Time on}	        & \colhead{Synthesized}    &	\colhead{Phase}	&	\colhead{Flux} 	\\[-9pt]
\colhead{Object}   & \colhead{Obs. Date}  & \colhead{Source}	   & \colhead{Beam}			&	\colhead{Calib.}	&	\colhead{Calib.}   \\[-9pt]
\colhead{}	            & \colhead{}              & \colhead{(hh:mm:ss)}	    & \colhead{($\arcsec \times \arcsec$)} &	\colhead{}	&	\colhead{} 	
}	
\startdata																				
GJ 564 BC	            &	2019 Jan 21	&	01:34:10	& 4.17 $\times$ 3.59 & J1443$+$2501 & 3C286 \\
		            	&	2019 Feb 05	&	01:33:59	& 3.15 $\times$ 2.55 & J1443$+$2501 & 3C286 \\
J2140+16          &	2019 Feb 03 &	04:15:16	& 4.10 $\times$ 3.65 & J2139$+$1423 & 3C286 \\
LP 415$-$20           	&	2018 Dec 28 &	04:15:30	& 4.09 $\times$ 3.71 & J0431$+$2037 & 3C147 \\
\enddata
\tablecomments{All observations were at 4--8 GHz and taken during C configuration at the VLA.}
\end{deluxetable}

%% file: TABLE_imaging.tex
\setlength{\tabcolsep}{0.05in}
\begin{deluxetable*}{lccccccccl}[htp]
\tabletypesize{\scriptsize}
\tablecaption{Flux density measurements \label{table:imaging}}
\tablehead{	
\colhead{}	        & 
\colhead{Stokes I}	& 
\colhead{ }      	& 
\colhead{ }         & 
\colhead{}          & 
\colhead{Stokes V}  & 
\colhead{}	        &	
\colhead{}	        & 
\colhead{}  \\[-9pt]
\colhead{}	                                & 
\colhead{Peak Brightness } 					&
\colhead{Peak Brightness } 					&
\colhead{Integrated $F_{\nu}$}              & 
\colhead{$[L_{\nu}]$\tablenotemark{\tiny{a}}} & 
\colhead{$F_{\nu}$}                         & 
\colhead{$[L_{\nu}]$\tablenotemark{\tiny{a}}} & 
\colhead{\% Circ.}	                        & 
\colhead{Notes}  \\[-9pt]
\colhead{Object}                            & 
\colhead{uvmodelfit}                        & 
\colhead{imfit}                             & 
\colhead{imfit}                             & 
\colhead{}	                                & 
\colhead{}                                  & 
\colhead{}                                  &	
\colhead{Poln.\tablenotemark{\tiny{a}}}	    & 
\colhead{} \\[-9pt]
\colhead{}	                        & 
\colhead{$(\mu$Jy beam$^{-1}$)}                 & 
\colhead{$(\mu$Jy beam$^{-1}$)}                 & 
\colhead{$(\mu$Jy)}                 & 
\colhead{$\log[$(erg s$^{-1}$ Hz$^{-1}$)$]$}	&
\colhead{$(\mu$Jy)}                 & 
\colhead{$\log[$(erg s$^{-1}$ Hz$^{-1}$)$]$}	&	
\colhead{}                          & 
\colhead{} 
}																					
\startdata																				
GJ 564 BC   & 22.7$\pm$2.7 	& 19.4$\pm$4.5  & 21.6$\pm$8.7  & 12.9  & $<$12.0   & $<$12.7  &  $\leq$58.8$_{-19.6}^{+24.0}$   & Epoch: 2019 Jan 21 \\[4pt] 
            & 31.3$\pm$2.2  & 30.1$\pm$3.7  & 38.7$\pm$7.6  & 13.1  & $<$9.3    & $<$12.6  &  $\leq$30.4$_{-9.8}^{+12.7}$    & Epoch: 2019 Feb 05 \\[4pt] 
J2140+1625  & 15.6$\pm$1.7  & 20.1$\pm$3.3  & 16.0$\pm$5.0  & 13.4  & $<$9.0    & $<$13.4  &  $\leq$43.6$_{-13.8}^{+20.8}$   &   \\[4pt] 
LP 415$-$20	& 24.1$\pm$1.4  & 22.9$\pm$2.5  & 24.9$\pm$4.6  & 13.6  & $<$6.5    & $<$13.1  &  $\leq$28.0$_{-9.1}^{+11.2}$    &	 \\[4pt] 
\enddata
\tablecomments{We measured Stokes I flux densities in two ways: (1) fitting the cleaned image with the CASA task \texttt{imfit} and (2) fitting the uv visibilities with the task  \texttt{uvmodelfit} after subtracting other sources in the primary beam.   \texttt{imfit} returns both peak brightness and integrated flux density.  For point sources, the peak flux density should be consistent with the integrated flux density.   \texttt{uvmodelfit} returns formal errors that  are underestimated by at least a factor $\sqrt{\chi^2_{\text{R}}}$, which we have corrected for in the reported errors.   No Stokes V sources were detected, so we report the 3$\sigma_{\text{rms}}$ upper  limit.}
\tablenotetext{a}{Calculated using the peak $F_{\nu}$ fitted with \texttt{imfit}. Uncertainties in log-luminosity are $<$0.1 and do not affect the presented analysis.}
\end{deluxetable*}

%% file: SECTION_discussion.tex
\section{Discussion}\label{sec.Discussion}
\begin{figure}[!ht]
\epsscale{1.18}
\plotone{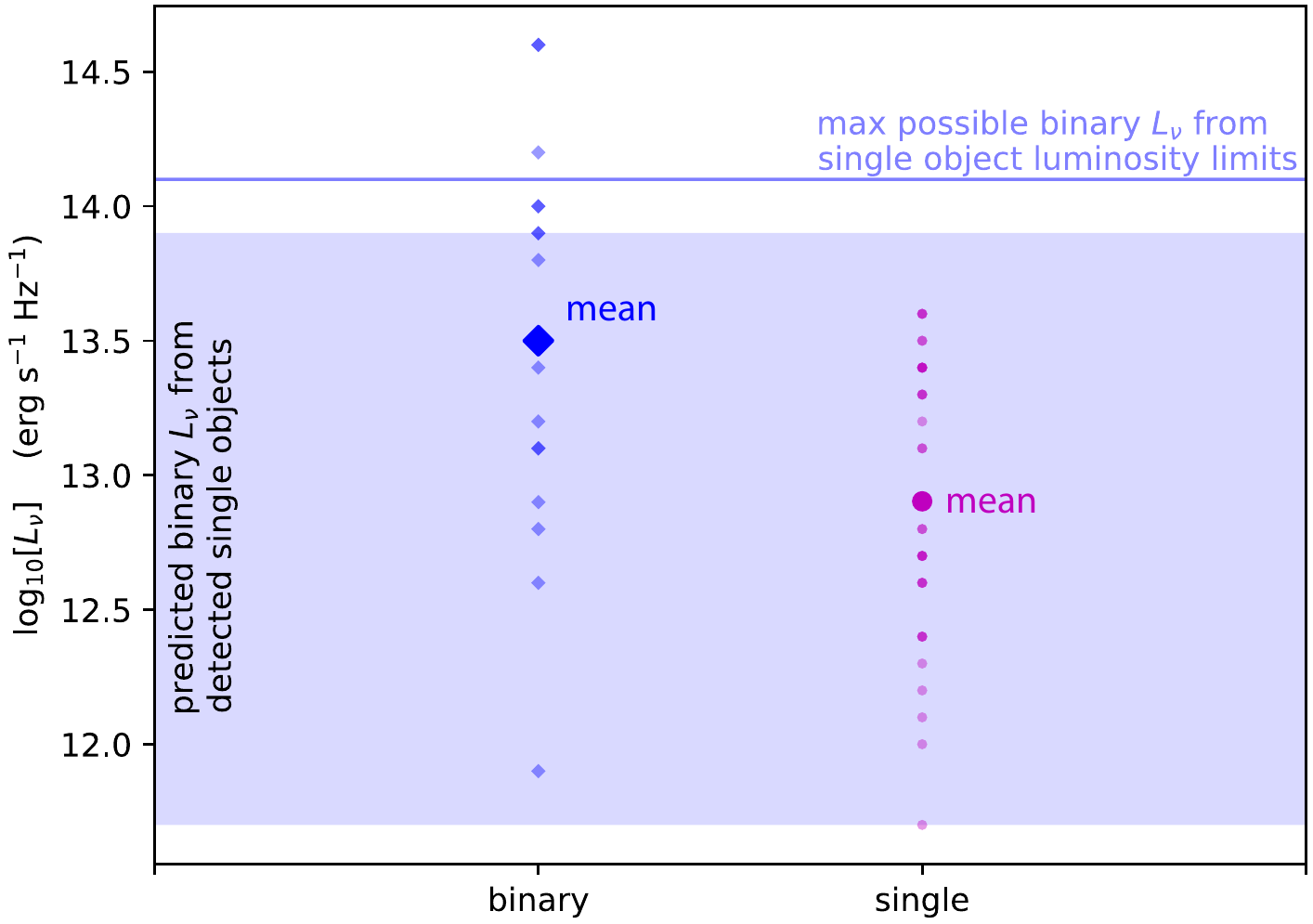}
\caption{\label{fig:binaryLum} 
Radio luminosities of detected quiescent radio emission from ultracool dwarfs (translucent small markers) and mean values for binary versus single objects (solid big markers).  Uncertainties are less than marker sizes. Shaded region corresponds to predicted binary luminosity ranges if individual binary components follow the same luminosity distribution as single objects. 
Detected binary ultracool dwarf systems are on average more radio luminous than single objects.
We use the compilations of single objects in \citet{kao2020a}, which includes 82 ultracool M dwarfs, 74 L dwarfs, and 23 T/Y dwarfs }
\end{figure}

\input{TABLE_quiescent}

We have detected radio emission from each of our three targets. In comparison, volume-limited radio surveys yield detection rates between $\sim$5--10\% for M, L, and T ultracool dwarfs \citep{Antonova2013AA...549A.131A, Lynch2016MNRAS.457.1224L, RouteWolszczan2016ApJ...830...85R}. The high success rate of our small sample suggests that binary ultracool dwarf systems with at least one rapidly rotating component ($v \sin i \gtrsim 35$ km s$^{-1}$) may be promising candidates for radio studies. 

Our high detection rate is subject to small-number statistics, and detection rates in general do not account for systematics like observational sensitivity or objects' intrinsic magnetospheric luminosity. In a forthcoming paper, we modify the occurrence rate framework developed by \citet{kao2020a} for single-object systems to allow for direct comparisons of radio occurrence rates between binary and single-object systems.
 
Another possible contributing factor for our high detection rate is that binaries may be intrinsically brighter at radio frequencies than single-object systems. This is because binaries have twice as many components that can produce magnetospheric radio emission than single-object systems do. Indeed, we compiled all available measurements of detected quiescent radio emission from binary ultracool dwarfs in the literature in Table \ref{table:quiescent} and show in Figure \ref{fig:binaryLum} that their mean quiescent radio luminosities are brighter than detected single objects. 

Intriguingly, the brightest binary systems may exceed luminosities predicted from single objects. We expect that binary luminosities will not exceed two times the maximum luminosity of single objects if binarity does not affect the luminosities of individual binary components.  In Figure \ref{fig:binaryLum}, we show that the brightest detected single systems cannot account for the high luminosities of the brightest detected binaries. This figure accounts for an individual object's intrinsic variability by including measurements from repeated observations of detected objects.

One immediate implication is that binary systems at farther distances than single objects may be detectable, increasing the number of observationally accessible systems. For instance, our targets 2MASS J21402931+1625183 and LP 415-20 are the most distant radio-bright ultracool dwarfs that have been detected at GHz frequencies to date, with distances of $33.1 \pm 0.5$ and $39.6884 \pm 0.8060$ pc, respectively. Before this work, the most distant radio-bright ultracool dwarf system was the binary 2MASS J09522188-1924319 ($29.0 \pm 0.13$ pc) \citep{McLean2012ApJ...746...23M, Gaia2018yCat.1345....0G}.  For comparison, the most distant detected single object is the L8.5 dwarf 2MASS J10430758+2225236 \citep{Kao2016ApJ...818...24K, Kao2018ApJS..237...25K} at $16.4\pm0.2$pc \citep{Schmidt2010AJ....139.1808S}.

\begin{figure}[!ht]
\epsscale{1.18}
\plotone{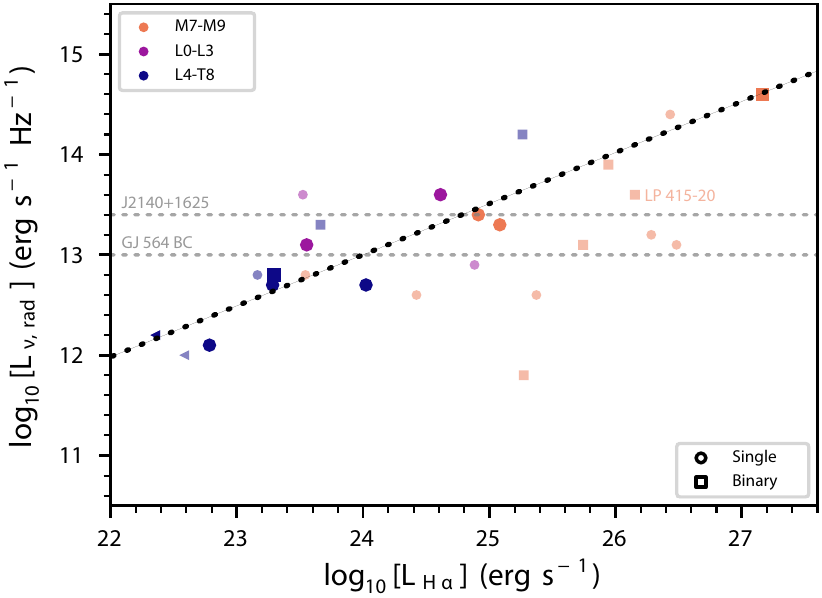}
\caption{\label{fig:halpha} 
Radio versus H$\alpha$ luminosities for binary and single-object systems with detected quiescent radio emission for systems that are confirmed auroral emitters (solid) versus those that are not (translucent), depicting the correlation between quiescent radio and H$\alpha$ luminosities (black dotted line) by \citet{Pineda2017ApJ...846...75P}. Only one of our observed targets, LP~415-20, has a non-zero or existing H$\alpha$ measurement.  Grey dashed lines correspond to the measured radio luminosities for other two targets. Uncertainties are comparable to or smaller than the symbol sizes, except for binaries which cannot be resolved.  For these systems, we consider total emission as coming from a single object with bolometric luminosity consistent with a single object that has an average spectral type of pair, which will produce additional systematics at the $\sim$0.3 dex level.   These systematics are not relevant to our science discussion, so we do not show them  here.}
\end{figure}

In Figure \ref{fig:halpha}, we also compare the quiescent radio luminosities of our targets to their H$\alpha$ luminosities, which \citet{Pineda2017ApJ...846...75P} showed correlate for aurorae-emitting ultracool dwarfs. We convert H$\alpha$ equivalent widths to $L_{\mathrm{H}\alpha}/L_{\mathrm{bol}}$ using the $\chi$ values from \citet{Schmidt2014PASP..126..642S}. For systems without measured $L_{\mathrm{bol}}$ by \citet{Dupuy2017ApJS..231...15D}, we use the \citet{Filipazzo2015ApJ...810..158F} relations between spectral type and $L_{\mathrm{bol}}$.

For LP~415-20, we do not detect any circularly polarized radio pulses that could indicate the presence of aurorae.  While its H$\alpha$ and quiescent radio luminosities cannot conclusively rule in or rule out the possibility that this system may exhibit radio aurorae in follow-up observations, we note that its strong H$\alpha$ emission is slightly rightward of the proposed auroral correlation and is similar to other late M dwarf systems that do not show periodic radio pulses. 

For 2MASS~J21402931+1625183, we do not observe radio pulses in our data. If this system later proves to be auroral, its quiescent radio luminosity predicts an H$\alpha$ luminosity of $[L_{\mathrm{H}\alpha}] \approx 24.8$ $[\mathrm{erg\,s}^{-1}]$ though scatter likely exists along the proposed auroral radio-H$\alpha$ correlation.  This translates to an EW of 0.7 \AA\, if the emission originates from the M8 dwarf or an EW of 3 \AA\, for the L0.5 dwarf. \citet{Gizis2000AJ....120.1085G} reported an H$\alpha$ EW of 0 \AA, but they note that their H$\alpha$ measurements should be taken with caution due to observing challenges. In particular, their data suggests that they may have had difficulty detecting H$\alpha$ emission at the 0.7 \AA\, level, 
  since they measure a non-zero EW $<$1 \AA\, for only one target in their sample of 60 late M and early L dwarfs yet most objects in this spectral type range exhibit H$\alpha$ emission.  Additionally,  because their observations do not resolve individual binary components, H$\alpha$ emission on the faster rotating L dwarf may be obscured by the continuum of the brighter M8 dwarf. If this is the case and it exhibited weak radio pulses below our detection threshold, then resolved spectroscopy of the binary may yield a confident H$\alpha$ detection.

Finally, GJ 564 BC does not have an available H$\alpha$ measurement in the literature though we detect a tentative radio burst that may indicate aurorae. If its tentative radio burst is indeed auroral in nature, its averaged quiescent radio flux predicts an H$\alpha$ luminosity of $[L_{\mathrm{H}\alpha}] \approx 24.0$ $[\mathrm{erg\,s}^{-1}]$, which translates to an EW of 2.6 \AA.

The radio observations presented here and the growing population of known radio-bright ultracool dwarf binaries will also be valuable benchmarks for  calibrating magnetic models in the substellar mass regime.  For instance,  \citet{Mullan2010ApJ...713.1249M} found that non-magnetic models cannot replicate both the observed luminosity and $T_{\mathrm{eff}}$ of GJ~564 BC.  The addition of a strong internal magnetic field corresponding to a $\gtrsim$400 G surface field strength impedes the onset of convection and helps resolve this issue.  If follow-up observations confirm that the tentative pulse that we observe from  GJ~564 BC is indeed auroral in nature, its surface-averaged magnetic field may be at least as strong as $\geq$1.3 kG \citep{Kao2016ApJ...818...24K}.

Additional detections and confirmations of the circularly polarized auroral pulses that usually coexist with ultracool dwarf quiescent radio emission \citep[e.g.,][]{Berger2001Natur.410..338B, Hallinan2007ApJ...663L..25H, Hallinan2008ApJ...684..644H, Williams2015ApJ...808..189W, Kao2016ApJ...818...24K, Kao2018ApJS..237...25K} will yield more direct tests to calibrate such magnetism-dependent models.

%% file: TABLE_quiescent.tex
\setlength{\tabcolsep}{0.05in}
\begin{deluxetable*}{lllr@{\hspace{0.01in}}c@{\hspace{0.01in}}lr@{\hspace{0.01in}}c@{\hspace{0.01in}}lcr@{\hspace{0.01in}}c@{\hspace{0.01in}}lcl}[!h]
\tabletypesize{\scriptsize}
\tablecaption{Specific luminosities for detected quiescent radio emission in binary ultracool dwarf systems\label{table:quiescent}}
\tablehead{																									\colhead{Object Name}			&	
\colhead{Other Name}			&	
\colhead{SpT}					&	
\multicolumn{3}{c}{$\pi$}		&	
\multicolumn{3}{c}{$d$}			&	
\colhead{ref}					&	
\multicolumn{3}{c}{$F_{\nu}$}	&	
\colhead{$[L_{\nu}]$\tablenotemark{\tiny{a}}} & 
\colhead{ref}	\\[-9pt]
\colhead{}						&	
\colhead{}						&	
\colhead{}						&	
\multicolumn{3}{c}{(mas)}		&	
\multicolumn{3}{c}{(pc)}		&	
\colhead{}						&	
\multicolumn{3}{c}{($\mu$Jy)}	&	
\colhead{$\log[$(erg s$^{-1}$ Hz$^{-1}$)$]$}	&
\colhead{}
}																	
\startdata	
2MASS J00043484$-$4044058 	&	GJ 1001 B					&	L5+L5		&	82.0946	&	$\pm$	&	0.3768	&	12.1811	&	$\pm$	&	0.0559	&	30	&	100.0	&	$\pm$	&	8.3	&	13.2	&	20	\\
2MASS J00275592+2219328		&	LP349$-$25					&	M7+M8		&	69.2	&	$\pm$	&	0.9		&	14.5	&	$\pm$	&	0.2		&	13	&	262		&	$\pm$	&	40	&	13.8	&	23	\\
							&								&				&			&			&			&			&			&			&		&	320		&	$\pm$	&	21	&	13.9	&	23	\\
							&								&				&			&			&			&			&			&			&		&	338		&	$\pm$	&	54	&	13.9	&	23	\\
							&								&				&			&			&			&			&			&			&		&	365		&	$\pm$	&	16	&	14.0	&	24	\\
							&								&				&			&			&			&			&			&			&		&	383		&	$\pm$	&	27	&	14.0	&	23	\\
2MASS J04234858$-$0414035		&	SDSS J042348.57$-$041403.5	&	L6.5+T2		&	67.8584	&	$\pm$	&	1.5052	&	14.7366	&	$\pm$	&	0.3269	&	30	&	15.4	&	$\pm$	&	1.2	&	12.6	&	19	\\
							&								&				&			&			&			&			&			&			&		&	26.7	&	$\pm$	&	3.1	&	12.8	&	18	\\
WISE\phm{S} J072003.20$-$084651.2AB	&								&	M9+T5		&	142		&	$\pm$	&	38		&	7		&	$\pm$	&	2		&	26	&	15		&	$\pm$	&	3	&	11.9	&	10	\\
2MASS J12560183$-$1257276a	&	VHS J125601.92$-$125723.9a	&	M7.5+M7.5	&	78.8	&	$\pm$	&	6.4		&	12.7	&	$\pm$	&	1.0		&	15	&	60		&	$\pm$	&	3	&	13.1	&	16	\\
2MASS J13153094$-$2649513AB	&								&	L3.5+T7		&	53.8729	&	$\pm$	&	1.1265	&	18.5622	&	$\pm$	&	0.3881	&	30	&	370		&	$\pm$	&	50	&	14.2	&	9	\\
2MASS J13142039+1320011		&	NLTT 33370					&	M7.0+M7.0	&	57.9750	&	$\pm$	&	0.0450	&	17.2488	&	$\pm$	&	0.0134	&	12	&	1099	&	$\pm$	&	18	&	14.6	&	21	\\
							&								&				&			&			&			&			&			&			&		&	1032	&	$\pm$	&	16	&	14.6	&	21	\\
GJ 564 BC					&	HD 130948B					&	L4+L4		&	54.9068	&	$\pm$	&	0.0684	&	18.2127	&	$\pm$	&	0.0227	&	30	&	19.4	&	$\pm$	&	4.5	&	12.9	&	1	\\
							&								&				&			&			&			&			&			&   		&		&	30.1	&	$\pm$	&	3.7	&	13.1	&	1	\\
2MASS J21402931+1625183		&								&	M8+L0.5		&	30.1972	&	$\pm$	&	0.4340	&	33.1157	&	$\pm$	&	0.4759	&	30	&	20.1	&	$\pm$	&	3.3	&	13.4	&	1	\\
\enddata
\tablecomments{This table does not include the M7+M7 binary 2MASS J09522188-1924319 due to possibility that the radio emission detected from this object may be flaring rather than quiescent emission. \citep{McLean2012ApJ...746...23M} detected $233 \pm 15$ $mu$Jy emission from this object, which corresponds to $[L_{\nu}] = 14.4$. However, they reported that followup observations at 4.96 and 8.46 GHz after the initial detection did not yield a detection to a limit of 69 $\mu$Jy, or a factor of 2.4 below the original detection.  They concluded that the initial detection was likely a flare or that 2MASS J09522188-1924319 exhibits long-term variability.  We also exclude LP 415-20 because of compelling evidence that the primary component of this system may in fact be unresolved ultracool dwarf binaries \citep{Dupuy2017ApJS..231...15D}.  }
\tablenotetext{a}{Uncertainties in log-luminosity are $<$0.1 and do not affect the presented analysis.}
\tablenotetext{b}{\citet{McLean2012ApJ...746...23M}, \citet{Forbrich2016ApJ...827...22F}, and \citet{Williams2015ApJ...799..192W} also observe 2MASS J13142039+1320011 at radio frequencies but do not separately report quiescent emission.}
\tablerefs{			
(1)	    This paper	;
(2)		\citet{Berger2001Natur.410..338B}	;
(3)		\citet{Berger2002ApJ...572..503B}	;
(4)		\citet{Berger2005ApJ...627..960B}	;
(5)		\citet{Berger2006ApJ...648..629B}	;
(6)		\citet{Berger2008ApJ...673.1080B}	;
(7)		\citet{Berger2008ApJ...676.1307B}	;
(8)		\citet{Berger2009ApJ...695..310B}	;
(9)		\citet{Burgasser2013ApJ...762L...3B}	;
(10)		\citet{Burgasser2015AJ....150..180B}	;
(11)		\citet{BurgasserPutnam2005ApJ...626..486B}	;
(12)		\citet{Dupuy2016ApJ...827...23D}	;
(13)		\citet{Dupuy2017ApJS..231...15D}	;
(14)		\citet{Faherty2012ApJ...752...56F}	;
(15)		\citet{Gauza2015ApJ...804...96G}	;
(16)		\citet{Guirado2018AA...610A..23G}	;
(17)		\citet{Hallinan2006ApJ...653..690H}	;
(18)		\citet{Kao2016ApJ...818...24K}	;
(19)		\citet{Kao2018ApJS..237...25K}	;
(20)		\citet{Lynch2016MNRAS.457.1224L}	;
(21)		\citet{McLean2011ApJ...741...27M}	;
(22)		\citet{Osten2006ApJ...637..518O}	;
(23)		\citet{Osten2009ApJ...700.1750O}	;
(24)		\citet{Phan-Bao2007ApJ...658..553P}	;
(25)		\citet{Richey-Yowell2020}	;
(26)		\citet{Scholz2014AA...561A.113S}	;
(27)		\citet{Schmidt2010AJ....139.1808S}	;
(28)		\citet{Williams2013ApJ...767L..30W}	;
(29)		\citet{Williams2015ApJ...808..189W}	;
(30)		\citet{Gaia2018yCat.1345....0G}	}
\end{deluxetable*}

%% file: SECTION_conclusions.tex
\section{Conclusions}\label{sec.Conclusion}

Binary ultracool dwarf systems are promising targets for radio studies of their magnetic activity due to the relative ease of constraining the masses and ages of their individual components compared to single objects. Here, we show that they may also be promising targets because the quiescent radio emission of the brightest binaries may be overly luminous compared to naively pairing single-object systems.

We present new radio observations and detections of quiescent radio emission from three binary systems that were previously not known to be radio emitters. Our detections of 2MASS J21402931+1625183 and LP 415-20 represent the farthest known radio-bright ultracool dwarfs sans coherent radio emission.  We also tentatively detect possible 4--6 GHz aurorae from GJ 564 BC. If follow-up monitoring shows that our targets are auroral in nature, we predict their H$\alpha$  luminosities based on the proposed radio-H$\alpha$ correlation by \citet{Pineda2017ApJ...846...75P}.  The three new detections that we present increase the number of known radio ultracool dwarf systems to 25. Existing radio detections of ultracool dwarf systems suggest that rapidly rotating objects with $v\sin i \gtrsim 35$ km s$^{-1}$ may be promising targets for radio studies.  Our observations support this picture.

%% file: manuscript.bbl
\begin{thebibliography}{}
\expandafter\ifx\csname natexlab\endcsname\relax\def\natexlab#1{#1}\fi
\providecommand{\url}[1]{\href{#1}{#1}}
\providecommand{\dodoi}[1]{doi:~\href{http://doi.org/#1}{\nolinkurl{#1}}}
\providecommand{\doeprint}[1]{\href{http://ascl.net/#1}{\nolinkurl{http://ascl.net/#1}}}
\providecommand{\doarXiv}[1]{\href{https://arxiv.org/abs/#1}{\nolinkurl{https://arxiv.org/abs/#1}}}

\bibitem[{{Antonova} {et~al.}(2013){Antonova}, {Hallinan}, {Doyle}, {Yu},
  {Kuznetsov}, {Metodieva}, {Golden}, \& {Cruz}}]{Antonova2013AA...549A.131A}
{Antonova}, A., {Hallinan}, G., {Doyle}, J.~G., {et~al.} 2013, \aap, 549, A131,
  \dodoi{10.1051/0004-6361/201118583}

\bibitem[{{Berger}(2002)}]{Berger2002ApJ...572..503B}
{Berger}, E. 2002, \apj, 572, 503, \dodoi{10.1086/340301}

\bibitem[{{Berger}(2006)}]{Berger2006ApJ...648..629B}
---. 2006, \apj, 648, 629, \dodoi{10.1086/505787}

\bibitem[{{Berger} {et~al.}(2001){Berger}, {Ball}, {Becker}, {Clarke}, {Frail},
  {Fukuda}, {Hoffman}, {Mellon}, {Momjian}, {Murphy}, {Teng}, {Woodruff},
  {Zauderer}, \& {Zavala}}]{Berger2001Natur.410..338B}
{Berger}, E., {Ball}, S., {Becker}, K.~M., {et~al.} 2001, \nat, 410, 338

\bibitem[{{Berger} {et~al.}(2005){Berger}, {Rutledge}, {Reid}, {Bildsten},
  {Gizis}, {Liebert}, {Mart{\'{\i}}n}, {Basri}, {Jayawardhana}, {Brandeker},
  {Fleming}, {Johns-Krull}, {Giampapa}, {Hawley}, \&
  {Schmitt}}]{Berger2005ApJ...627..960B}
{Berger}, E., {Rutledge}, R.~E., {Reid}, I.~N., {et~al.} 2005, \apj, 627, 960,
  \dodoi{10.1086/430343}

\bibitem[{{Berger} {et~al.}(2008{\natexlab{a}}){Berger}, {Basri}, {Gizis},
  {Giampapa}, {Rutledge}, {Liebert}, {Mart{\'{\i}}n}, {Fleming}, {Johns-Krull},
  {Phan-Bao}, \& {Sherry}}]{Berger2008ApJ...676.1307B}
{Berger}, E., {Basri}, G., {Gizis}, J.~E., {et~al.} 2008{\natexlab{a}}, \apj,
  676, 1307, \dodoi{10.1086/529131}

\bibitem[{{Berger} {et~al.}(2008{\natexlab{b}}){Berger}, {Gizis}, {Giampapa},
  {Rutledge}, {Liebert}, {Mart{\'\i}n}, {Basri}, {Fleming}, {Johns-Krull},
  {Phan-Bao}, \& {Sherry}}]{Berger2008ApJ...673.1080B}
{Berger}, E., {Gizis}, J.~E., {Giampapa}, M.~S., {et~al.} 2008{\natexlab{b}},
  \apj, 673, 1080, \dodoi{10.1086/524769}

\bibitem[{{Berger} {et~al.}(2009){Berger}, {Rutledge}, {Phan-Bao}, {Basri},
  {Giampapa}, {Gizis}, {Liebert}, {Mart{\'{\i}}n}, \&
  {Fleming}}]{Berger2009ApJ...695..310B}
{Berger}, E., {Rutledge}, R.~E., {Phan-Bao}, N., {et~al.} 2009, \apj, 695, 310,
  \dodoi{10.1088/0004-637X/695/1/310}

\bibitem[{{Bolton} {et~al.}(2004){Bolton}, {Thorne}, {Bourdarie}, {de Pater},
  \& {Mauk}}]{bolton2004}
{Bolton}, S.~J., {Thorne}, R.~M., {Bourdarie}, S., {de Pater}, I., \& {Mauk},
  B. 2004, {Jupiter's inner radiation belts}, ed. F.~{Bagenal}, T.~E.
  {Dowling}, \& W.~B. {McKinnon}, 671--688

\bibitem[{{Brandt} \& {Huang}(2015)}]{Brandt2015ApJ...807...58B}
{Brandt}, T.~D., \& {Huang}, C.~X. 2015, \apj, 807, 58,
  \dodoi{10.1088/0004-637X/807/1/58}

\bibitem[{{Burgasser} {et~al.}(2015){Burgasser}, {Melis}, {Todd}, {Gelino},
  {Hallinan}, \& {Bardalez Gagliuffi}}]{Burgasser2015AJ....150..180B}
{Burgasser}, A.~J., {Melis}, C., {Todd}, J., {et~al.} 2015, \aj, 150, 180,
  \dodoi{10.1088/0004-6256/150/6/180}

\bibitem[{{Burgasser} {et~al.}(2013){Burgasser}, {Melis}, {Zauderer}, \&
  {Berger}}]{Burgasser2013ApJ...762L...3B}
{Burgasser}, A.~J., {Melis}, C., {Zauderer}, B.~A., \& {Berger}, E. 2013,
  \apjl, 762, L3, \dodoi{10.1088/2041-8205/762/1/L3}

\bibitem[{{Burgasser} \& {Putman}(2005)}]{BurgasserPutnam2005ApJ...626..486B}
{Burgasser}, A.~J., \& {Putman}, M.~E. 2005, \apj, 626, 486,
  \dodoi{10.1086/429788}

\bibitem[{{Clarke} {et~al.}(2004){Clarke}, {Grodent}, {Cowley}, {Bunce},
  {Zarka}, {Connerney}, \& {Satoh}}]{Clarke2004jpsm.book..639C}
{Clarke}, J.~T., {Grodent}, D., {Cowley}, S.~W.~H., {et~al.} 2004, {Jupiter's
  aurora}, ed. F.~{Bagenal}, T.~E. {Dowling}, \& W.~B. {McKinnon}, 639--670

\bibitem[{{Close} {et~al.}(2003){Close}, {Siegler}, {Freed}, \&
  {Biller}}]{Close2003ApJ...587..407C}
{Close}, L.~M., {Siegler}, N., {Freed}, M., \& {Biller}, B. 2003, \apj, 587,
  407, \dodoi{10.1086/368177}

\bibitem[{{Close} {et~al.}(2002){Close}, {Siegler}, {Potter}, {Brand ner}, \&
  {Liebert}}]{Close2002ApJ...567L..53C}
{Close}, L.~M., {Siegler}, N., {Potter}, D., {Brand ner}, W., \& {Liebert}, J.
  2002, \apjl, 567, L53, \dodoi{10.1086/339795}

\bibitem[{{Dupuy} {et~al.}(2016){Dupuy}, {Forbrich}, {Rizzuto}, {Mann},
  {Aller}, {Liu}, {Kraus}, \& {Berger}}]{Dupuy2016ApJ...827...23D}
{Dupuy}, T.~J., {Forbrich}, J., {Rizzuto}, A., {et~al.} 2016, \apj, 827, 23,
  \dodoi{10.3847/0004-637X/827/1/23}

\bibitem[{{Dupuy} \& {Liu}(2017)}]{Dupuy2017ApJS..231...15D}
{Dupuy}, T.~J., \& {Liu}, M.~C. 2017, \apjs, 231, 15,
  \dodoi{10.3847/1538-4365/aa5e4c}

\bibitem[{{Dupuy} {et~al.}(2009){Dupuy}, {Liu}, \&
  {Ireland}}]{Dupuy2009ApJ...692..729D}
{Dupuy}, T.~J., {Liu}, M.~C., \& {Ireland}, M.~J. 2009, \apj, 692, 729,
  \dodoi{10.1088/0004-637X/692/1/729}

\bibitem[{{Dupuy} {et~al.}(2019){Dupuy}, {Liu}, {Best}, {Mann}, {Tucker},
  {Zhang}, {Baraffe}, {Chabrier}, {Forveille}, {Metchev}, {Tremblin}, {Do},
  {Payne}, {Shappee}, {Bond}, {Cetre}, {Chun}, {Delorme}, {Jovanovic},
  {Lilley}, {Mawet}, {Ragland }, {Wetherell}, \&
  {Wizinowich}}]{Dupuy2019AJ....158..174D}
{Dupuy}, T.~J., {Liu}, M.~C., {Best}, W. M.~J., {et~al.} 2019, \aj, 158, 174,
  \dodoi{10.3847/1538-3881/ab3cd1}

\bibitem[{{Faherty} {et~al.}(2009){Faherty}, {Burgasser}, {Cruz}, {Shara},
  {Walter}, \& {Gelino}}]{Faherty2009AJ....137....1F}
{Faherty}, J.~K., {Burgasser}, A.~J., {Cruz}, K.~L., {et~al.} 2009, \aj, 137,
  1, \dodoi{10.1088/0004-6256/137/1/1}

\bibitem[{{Faherty} {et~al.}(2012){Faherty}, {Burgasser}, {Walter}, {Van der
  Bliek}, {Shara}, {Cruz}, {West}, {Vrba}, \&
  {Anglada-Escud{\'e}}}]{Faherty2012ApJ...752...56F}
{Faherty}, J.~K., {Burgasser}, A.~J., {Walter}, F.~M., {et~al.} 2012, \apj,
  752, 56, \dodoi{10.1088/0004-637X/752/1/56}

\bibitem[{{Filippazzo} {et~al.}(2015){Filippazzo}, {Rice}, {Faherty}, {Cruz},
  {Van Gordon}, \& {Looper}}]{Filipazzo2015ApJ...810..158F}
{Filippazzo}, J.~C., {Rice}, E.~L., {Faherty}, J., {et~al.} 2015, \apj, 810,
  158, \dodoi{10.1088/0004-637X/810/2/158}

\bibitem[{{Forbrich} {et~al.}(2016){Forbrich}, {Dupuy}, {Reid}, {Berger},
  {Rizzuto}, {Mann}, {Liu}, {Aller}, \& {Kraus}}]{Forbrich2016ApJ...827...22F}
{Forbrich}, J., {Dupuy}, T.~J., {Reid}, M.~J., {et~al.} 2016, \apj, 827, 22,
  \dodoi{10.3847/0004-637X/827/1/22}

\bibitem[{{Gagn{\'e}} {et~al.}(2018){Gagn{\'e}}, {Mamajek}, {Malo}, {Riedel},
  {Rodriguez}, {Lafreni{\`e}re}, {Faherty}, {Roy-Loubier}, {Pueyo}, {Robin}, \&
  {Doyon}}]{Gagne2018ApJ...856...23G}
{Gagn{\'e}}, J., {Mamajek}, E.~E., {Malo}, L., {et~al.} 2018, \apj, 856, 23,
  \dodoi{10.3847/1538-4357/aaae09}

\bibitem[{{Gaia Collaboration}(2018)}]{Gaia2018yCat.1345....0G}
{Gaia Collaboration}. 2018, VizieR Online Data Catalog, I/345

\bibitem[{{Gauza} {et~al.}(2015){Gauza}, {B{\'e}jar}, {P{\'e}rez-Garrido},
  {Zapatero Osorio}, {Lodieu}, {Rebolo}, {Pall{\'e}}, \&
  {Nowak}}]{Gauza2015ApJ...804...96G}
{Gauza}, B., {B{\'e}jar}, V. J.~S., {P{\'e}rez-Garrido}, A., {et~al.} 2015,
  \apj, 804, 96, \dodoi{10.1088/0004-637X/804/2/96}

\bibitem[{{Gizis} {et~al.}(2013){Gizis}, {Burgasser}, {Berger}, {Williams},
  {Vrba}, {Cruz}, \& {Metchev}}]{Gizis2013ApJ...779..172G}
{Gizis}, J.~E., {Burgasser}, A.~J., {Berger}, E., {et~al.} 2013, \apj, 779,
  172, \dodoi{10.1088/0004-637X/779/2/172}

\bibitem[{{Gizis} {et~al.}(2000){Gizis}, {Monet}, {Reid}, {Kirkpatrick},
  {Liebert}, \& {Williams}}]{Gizis2000AJ....120.1085G}
{Gizis}, J.~E., {Monet}, D.~G., {Reid}, I.~N., {et~al.} 2000, \aj, 120, 1085,
  \dodoi{10.1086/301456}

\bibitem[{{Gizis} {et~al.}(2016){Gizis}, {Williams}, {Burgasser}, {Libralato},
  {Nardiello}, {Piotto}, {Bedin}, {Berger}, \&
  {Paudel}}]{Gizis2016AJ....152..123G}
{Gizis}, J.~E., {Williams}, P. K.~G., {Burgasser}, A.~J., {et~al.} 2016, \aj,
  152, 123, \dodoi{10.3847/0004-6256/152/5/123}

\bibitem[{{Guirado} {et~al.}(2018){Guirado}, {Azulay}, {Gauza},
  {P{\'e}rez-Torres}, {Rebolo}, {Climent}, \& {Zapatero
  Osorio}}]{Guirado2018AA...610A..23G}
{Guirado}, J.~C., {Azulay}, R., {Gauza}, B., {et~al.} 2018, \aap, 610, A23,
  \dodoi{10.1051/0004-6361/201732130}

\bibitem[{{Hallinan} {et~al.}(2006){Hallinan}, {Antonova}, {Doyle}, {Bourke},
  {Brisken}, \& {Golden}}]{Hallinan2006ApJ...653..690H}
{Hallinan}, G., {Antonova}, A., {Doyle}, J.~G., {et~al.} 2006, \apj, 653, 690,
  \dodoi{10.1086/508678}

\bibitem[{{Hallinan} {et~al.}(2008){Hallinan}, {Antonova}, {Doyle}, {Bourke},
  {Lane}, \& {Golden}}]{Hallinan2008ApJ...684..644H}
---. 2008, \apj, 684, 644, \dodoi{10.1086/590360}

\bibitem[{{Hallinan} {et~al.}(2007){Hallinan}, {Bourke}, {Lane}, {Antonova},
  {Zavala}, {Brisken}, {Boyle}, {Vrba}, {Doyle}, \&
  {Golden}}]{Hallinan2007ApJ...663L..25H}
{Hallinan}, G., {Bourke}, S., {Lane}, C., {et~al.} 2007, \apjl, 663, L25,
  \dodoi{10.1086/519790}

\bibitem[{{Hallinan} {et~al.}(2015){Hallinan}, {Littlefair}, {Cotter},
  {Bourke}, {Harding}, {Pineda}, {Butler}, {Golden}, {Basri}, {Doyle}, {Kao},
  {Berdyugina}, {Kuznetsov}, {Rupen}, \&
  {Antonova}}]{Hallinan2015Natur.523..568H}
{Hallinan}, G., {Littlefair}, S.~P., {Cotter}, G., {et~al.} 2015, \nat, 523,
  568, \dodoi{10.1038/nature14619}

\bibitem[{{Harding} {et~al.}(2013){Harding}, {Hallinan}, {Konopacky},
  {Kratter}, {Boyle}, {Butler}, \& {Golden}}]{Harding2013A&A...554A.113H}
{Harding}, L.~K., {Hallinan}, G., {Konopacky}, Q.~M., {et~al.} 2013, \aap, 554,
  A113, \dodoi{10.1051/0004-6361/201220865}

\bibitem[{Horne {et~al.}(2008)Horne, Thorne, Glauert, Douglas~Menietti,
  Shprits, \& Gurnett}]{Horne2008}
Horne, R.~B., Thorne, R.~M., Glauert, S.~A., {et~al.} 2008, Nature Physics, 4,
  301 EP .
\newblock \url{http://dx.doi.org/10.1038/nphys897}

\bibitem[{{Hughes} {et~al.}(2021){Hughes}, {Boley}, {Osten}, {White}, \&
  {Leacock}}]{hughes2021AJ....162...43H}
{Hughes}, A.~G., {Boley}, A.~C., {Osten}, R.~A., {White}, J.~A., \& {Leacock},
  M. 2021, \aj, 162, 43, \dodoi{10.3847/1538-3881/ac02c3}

\bibitem[{{Hunter}(2007)}]{matplotlib}
{Hunter}, J.~D. 2007, Computing in Science and Engineering, 9, 90,
  \dodoi{10.1109/MCSE.2007.55}

\bibitem[{{Jackman} {et~al.}(2019){Jackman}, {Wheatley}, {Bayliss}, {Burleigh},
  {Casewell}, {Eigm{\"u}ller}, {Goad}, {Pollacco}, {Raynard}, {Watson}, \&
  {West}}]{Jackman2019MNRAS.485L.136J}
{Jackman}, J. A.~G., {Wheatley}, P.~J., {Bayliss}, D., {et~al.} 2019, \mnras,
  485, L136, \dodoi{10.1093/mnrasl/slz039}

\bibitem[{{Kao} {et~al.}(2019){Kao}, {Hallinan}, \&
  {Pineda}}]{Kao2019MNRAS.487.1994K}
{Kao}, M.~M., {Hallinan}, G., \& {Pineda}, J.~S. 2019, \mnras, 487, 1994,
  \dodoi{10.1093/mnras/stz1372}

\bibitem[{{Kao} {et~al.}(2016){Kao}, {Hallinan}, {Pineda}, {Escala},
  {Burgasser}, {Bourke}, \& {Stevenson}}]{Kao2016ApJ...818...24K}
{Kao}, M.~M., {Hallinan}, G., {Pineda}, J.~S., {et~al.} 2016, \apj, 818, 24,
  \dodoi{10.3847/0004-637X/818/1/24}

\bibitem[{{Kao} {et~al.}(2018){Kao}, {Hallinan}, {Pineda}, {Stevenson}, \&
  {Burgasser}}]{Kao2018ApJS..237...25K}
{Kao}, M.~M., {Hallinan}, G., {Pineda}, J.~S., {Stevenson}, D., \& {Burgasser},
  A. 2018, \apjs, 237, 25, \dodoi{10.3847/1538-4365/aac2d5}

\bibitem[{{Kao} \& {Shkolnik }(submitted)}]{kao2020a}
{Kao}, M.~M., \& {Shkolnik }, E. submitted, In prep.

\bibitem[{{Konopacky} {et~al.}(2010){Konopacky}, {Ghez}, {Barman}, {Rice},
  {Bailey}, {White}, {McLean}, \& {Duch{\^e}ne}}]{Konopacky2010ApJ...711.1087K}
{Konopacky}, Q.~M., {Ghez}, A.~M., {Barman}, T.~S., {et~al.} 2010, \apj, 711,
  1087, \dodoi{10.1088/0004-637X/711/2/1087}

\bibitem[{{Konopacky} {et~al.}(2012){Konopacky}, {Ghez}, {Fabrycky},
  {Macintosh}, {White}, {Barman}, {Rice}, {Hallinan}, \&
  {Duch{\^e}ne}}]{Konopacky2012ApJ...750...79K}
{Konopacky}, Q.~M., {Ghez}, A.~M., {Fabrycky}, D.~C., {et~al.} 2012, \apj, 750,
  79, \dodoi{10.1088/0004-637X/750/1/79}

\bibitem[{{Lynch} {et~al.}(2016){Lynch}, {Murphy}, {Ravi}, {Hobbs}, {Lo}, \&
  {Ward}}]{Lynch2016MNRAS.457.1224L}
{Lynch}, C., {Murphy}, T., {Ravi}, V., {et~al.} 2016, \mnras, 457, 1224,
  \dodoi{10.1093/mnras/stw050}

\bibitem[{{Martin} {et~al.}(2017){Martin}, {Mace}, {McLean}, {Logsdon}, {Rice},
  {Kirkpatrick}, {Burgasser}, {McGovern}, \&
  {Prato}}]{Martin2017ApJ...838...73M}
{Martin}, E.~C., {Mace}, G.~N., {McLean}, I.~S., {et~al.} 2017, \apj, 838, 73,
  \dodoi{10.3847/1538-4357/aa6338}

\bibitem[{{McLean} {et~al.}(2011){McLean}, {Berger}, {Irwin}, {Forbrich}, \&
  {Reiners}}]{McLean2011ApJ...741...27M}
{McLean}, M., {Berger}, E., {Irwin}, J., {Forbrich}, J., \& {Reiners}, A. 2011,
  \apj, 741, 27, \dodoi{10.1088/0004-637X/741/1/27}

\bibitem[{{McLean} {et~al.}(2012){McLean}, {Berger}, \&
  {Reiners}}]{McLean2012ApJ...746...23M}
{McLean}, M., {Berger}, E., \& {Reiners}, A. 2012, \apj, 746, 23,
  \dodoi{10.1088/0004-637X/746/1/23}

\bibitem[{{McMullin} {et~al.}(2007){McMullin}, {Waters}, {Schiebel}, {Young},
  \& {Golap}}]{CASA}
{McMullin}, J.~P., {Waters}, B., {Schiebel}, D., {Young}, W., \& {Golap}, K.
  2007, Astronomical Society of the Pacific Conference Series, Vol. 376, {CASA
  Architecture and Applications}, ed. R.~A. {Shaw}, F.~{Hill}, \& D.~J. {Bell},
  127

\bibitem[{{Miles-P{\'a}ez} {et~al.}(2017){Miles-P{\'a}ez}, {Pall{\'e}}, \&
  {Zapatero Osorio}}]{MilesPaez2017MNRAS.472.2297M}
{Miles-P{\'a}ez}, P.~A., {Pall{\'e}}, E., \& {Zapatero Osorio}, M.~R. 2017,
  \mnras, 472, 2297, \dodoi{10.1093/mnras/stx2191}

\bibitem[{{Mullan} \& {MacDonald}(2010)}]{Mullan2010ApJ...713.1249M}
{Mullan}, D.~J., \& {MacDonald}, J. 2010, \apj, 713, 1249,
  \dodoi{10.1088/0004-637X/713/2/1249}

\bibitem[{{Nichols} {et~al.}(2012){Nichols}, {Burleigh}, {Casewell}, {Cowley},
  {Wynn}, {Clarke}, \& {West}}]{Nichols2012ApJ...760...59N}
{Nichols}, J.~D., {Burleigh}, M.~R., {Casewell}, S.~L., {et~al.} 2012, \apj,
  760, 59, \dodoi{10.1088/0004-637X/760/1/59}

\bibitem[{{Osten} {et~al.}(2006){Osten}, {Hawley}, {Bastian}, \&
  {Reid}}]{Osten2006ApJ...637..518O}
{Osten}, R.~A., {Hawley}, S.~L., {Bastian}, T.~S., \& {Reid}, I.~N. 2006, \apj,
  637, 518, \dodoi{10.1086/498345}

\bibitem[{{Osten} {et~al.}(2009){Osten}, {Phan-Bao}, {Hawley}, {Reid}, \&
  {Ojha}}]{Osten2009ApJ...700.1750O}
{Osten}, R.~A., {Phan-Bao}, N., {Hawley}, S.~L., {Reid}, I.~N., \& {Ojha}, R.
  2009, \apj, 700, 1750, \dodoi{10.1088/0004-637X/700/2/1750}

\bibitem[{{Paudel} {et~al.}(2018){Paudel}, {Gizis}, {Mullan}, {Schmidt},
  {Burgasser}, {Williams}, \& {Berger}}]{Paudel2018ApJ...858...55P}
{Paudel}, R.~R., {Gizis}, J.~E., {Mullan}, D.~J., {et~al.} 2018, \apj, 858, 55,
  \dodoi{10.3847/1538-4357/aab8fe}

\bibitem[{{Paudel} {et~al.}(2020){Paudel}, {Gizis}, {Mullan}, {Williams},
  {Burgasser}, \& {Schmidt}}]{Paudel2020arXiv200410579P}
---. 2020, arXiv e-prints, arXiv:2004.10579.
\newblock \doarXiv{2004.10579}

\bibitem[{{Perley} {et~al.}(2011){Perley}, {Chandler}, {Butler}, \&
  {Wrobel}}]{Perley2011ApJ...739L...1P}
{Perley}, R.~A., {Chandler}, C.~J., {Butler}, B.~J., \& {Wrobel}, J.~M. 2011,
  \apjl, 739, L1, \dodoi{10.1088/2041-8205/739/1/L1}

\bibitem[{{Phan-Bao} {et~al.}(2007){Phan-Bao}, {Osten}, {Lim}, {Mart{\'\i}n},
  \& {Ho}}]{Phan-Bao2007ApJ...658..553P}
{Phan-Bao}, N., {Osten}, R.~A., {Lim}, J., {Mart{\'\i}n}, E.~L., \& {Ho}, P.
  T.~P. 2007, \apj, 658, 553, \dodoi{10.1086/511061}

\bibitem[{{Pineda} {et~al.}(2017){Pineda}, {Hallinan}, \&
  {Kao}}]{Pineda2017ApJ...846...75P}
{Pineda}, J.~S., {Hallinan}, G., \& {Kao}, M.~M. 2017, \apj, 846, 75,
  \dodoi{10.3847/1538-4357/aa8596}

\bibitem[{{Pineda} {et~al.}(2016){Pineda}, {Hallinan}, {Kirkpatrick}, {Cotter},
  {Kao}, \& {Mooley}}]{Pineda2016ApJ...826...73P}
{Pineda}, J.~S., {Hallinan}, G., {Kirkpatrick}, J.~D., {et~al.} 2016, \apj,
  826, 73, \dodoi{10.3847/0004-637X/826/1/73}

\bibitem[{{Potter} {et~al.}(2002){Potter}, {Mart{\'\i}n}, {Cushing}, {Baudoz},
  {Brandner}, {Guyon}, \& {Neuh{\"a}user}}]{Potter2002ApJ...567L.133P}
{Potter}, D., {Mart{\'\i}n}, E.~L., {Cushing}, M.~C., {et~al.} 2002, \apjl,
  567, L133, \dodoi{10.1086/339999}

\bibitem[{{Price-Whelan} {et~al.}(2018){Price-Whelan}, {Sip{\'{o}}cz},
  {G{\"u}nther}, {Lim}, {Crawford}, {Conseil}, {Shupe}, {Craig}, {Dencheva},
  {Ginsburg}, {VanderPlas}, {Bradley}, {P{'e}rez-Su{'a}rez}, {de Val-Borro},
  {Paper Contributors}, {Aldcroft}, {Cruz}, {Robitaille}, {Tollerud},
  {Coordination Committee}, {Ardelean}, {Babej}, {Bach}, {Bachetti}, {Bakanov},
  {Bamford}, {Barentsen}, {Barmby}, {Baumbach}, {Berry}, {Biscani}, {Boquien},
  {Bostroem}, {Bouma}, {Brammer}, {Bray}, {Breytenbach}, {Buddelmeijer},
  {Burke}, {Calderone}, {Cano Rodr{'i}guez}, {Cara}, {Cardoso}, {Cheedella},
  {Copin}, {Corrales}, {Crichton}, {D{ extquoteright}Avella}, {Deil},
  {Depagne}, {Dietrich}, {Donath}, {Droettboom}, {Earl}, {Erben}, {Fabbro},
  {Ferreira}, {Finethy}, {Fox}, {Garrison}, {Gibbons}, {Goldstein}, {Gommers},
  {Greco}, {Greenfield}, {Groener}, {Grollier}, {Hagen}, {Hirst}, {Homeier},
  {Horton}, {Hosseinzadeh}, {Hu}, {Hunkeler}, {Ivezi{'c}}, {Jain}, {Jenness},
  {Kanarek}, {Kendrew}, {Kern}, {Kerzendorf}, {Khvalko}, {King}, {Kirkby},
  {Kulkarni}, {Kumar}, {Lee}, {Lenz}, {Littlefair}, {Ma}, {Macleod},
  {Mastropietro}, {McCully}, {Montagnac}, {Morris}, {Mueller}, {Mumford},
  {Muna}, {Murphy}, {Nelson}, {Nguyen}, {Ninan}, {N{"o}the}, {Ogaz}, {Oh},
  {Parejko}, {Parley}, {Pascual}, {Patil}, {Patil}, {Plunkett}, {Prochaska},
  {Rastogi}, {Reddy Janga}, {Sabater}, {Sakurikar}, {Seifert}, {Sherbert},
  {Sherwood-Taylor}, {Shih}, {Sick}, {Silbiger}, {Singanamalla}, {Singer},
  {Sladen}, {Sooley}, {Sornarajah}, {Streicher}, {Teuben}, {Thomas},
  {Tremblay}, {Turner}, {Terr{'o}n}, {van Kerkwijk}, {de la Vega}, {Watkins},
  {Weaver}, {Whitmore}, {Woillez}, {Zabalza}, \& {Contributors}}]{astropy:2018}
{Price-Whelan}, A.~M., {Sip{\'{o}}cz}, B.~M., {G{\"u}nther}, H.~M., {et~al.}
  2018, AJ, 156, 123, \dodoi{10.3847/1538-3881/aabc4f}

\bibitem[{{Richey-Yowell} {et~al.}(2020){Richey-Yowell}, {Kao}, {Pineda},
  {Shkolnik}, \& {Hallinan}}]{Richey-Yowell2020}
{Richey-Yowell}, T., {Kao}, M.~M., {Pineda}, J.~S., {Shkolnik}, E.~L., \&
  {Hallinan}, G. 2020, \apj, 903, 74, \dodoi{10.3847/1538-4357/abb826}

\bibitem[{{Route} \& {Wolszczan}(2012)}]{RouteWolszczan2012ApJ...747L..22R}
{Route}, M., \& {Wolszczan}, A. 2012, \apjl, 747, L22,
  \dodoi{10.1088/2041-8205/747/2/L22}

\bibitem[{{Route} \&
  {Wolszczan}(2016{\natexlab{a}})}]{RouteWolszczan2016ApJ...821L..21R}
---. 2016{\natexlab{a}}, \apjl, 821, L21, \dodoi{10.3847/2041-8205/821/2/L21}

\bibitem[{{Route} \&
  {Wolszczan}(2016{\natexlab{b}})}]{RouteWolszczan2016ApJ...830...85R}
---. 2016{\natexlab{b}}, \apj, 830, 85, \dodoi{10.3847/0004-637X/830/2/85}

\bibitem[{{Sault} {et~al.}(1997){Sault}, {Oosterloo}, {Dulk}, \&
  {Leblanc}}]{sault1997}
{Sault}, R.~J., {Oosterloo}, T., {Dulk}, G.~A., \& {Leblanc}, Y. 1997, \aap,
  324, 1190

\bibitem[{{Schmidt} {et~al.}(2015){Schmidt}, {Hawley}, {West}, {Bochanski},
  {Davenport}, {Ge}, \& {Schneider}}]{Schmidt2015AJ....149..158S}
{Schmidt}, S.~J., {Hawley}, S.~L., {West}, A.~A., {et~al.} 2015, \aj, 149, 158,
  \dodoi{10.1088/0004-6256/149/5/158}

\bibitem[{{Schmidt} {et~al.}(2014){Schmidt}, {West}, {Bochanski}, {Hawley}, \&
  {Kielty}}]{Schmidt2014PASP..126..642S}
{Schmidt}, S.~J., {West}, A.~A., {Bochanski}, J.~J., {Hawley}, S.~L., \&
  {Kielty}, C. 2014, \pasp, 126, 642, \dodoi{10.1086/677403}

\bibitem[{{Schmidt} {et~al.}(2010){Schmidt}, {West}, {Hawley}, \&
  {Pineda}}]{Schmidt2010AJ....139.1808S}
{Schmidt}, S.~J., {West}, A.~A., {Hawley}, S.~L., \& {Pineda}, J.~S. 2010, \aj,
  139, 1808, \dodoi{10.1088/0004-6256/139/5/1808}

\bibitem[{{Scholz}(2014)}]{Scholz2014AA...561A.113S}
{Scholz}, R.~D. 2014, \aap, 561, A113, \dodoi{10.1051/0004-6361/201323015}

\bibitem[{{Shulyak} {et~al.}(2017){Shulyak}, {Reiners}, {Engeln}, {Malo},
  {Yadav}, {Morin}, \& {Kochukhov}}]{Shulyak2017NatAs...1E.184S}
{Shulyak}, D., {Reiners}, A., {Engeln}, A., {et~al.} 2017, Nature Astronomy, 1,
  0184, \dodoi{10.1038/s41550-017-0184}

\bibitem[{{Siegler} {et~al.}(2003){Siegler}, {Close}, {Mamajek}, \&
  {Freed}}]{Siegler2003ApJ...598.1265S}
{Siegler}, N., {Close}, L.~M., {Mamajek}, E.~E., \& {Freed}, M. 2003, \apj,
  598, 1265, \dodoi{10.1086/378935}

\bibitem[{{Treumann}(2006)}]{Treumann2006AARv..13..229T}
{Treumann}, R.~A. 2006, \aapr, 13, 229, \dodoi{10.1007/s00159-006-0001-y}

\bibitem[{{Turnpenney} {et~al.}(2017){Turnpenney}, {Nichols}, {Wynn}, \&
  {Casewell}}]{Turnpenney2017MNRAS.470.4274T}
{Turnpenney}, S., {Nichols}, J.~D., {Wynn}, G.~A., \& {Casewell}, S.~L. 2017,
  \mnras, 470, 4274, \dodoi{10.1093/mnras/stx1508}

\bibitem[{{van der Walt} {et~al.}(2011){van der Walt}, {Colbert}, \&
  {Varoquaux}}]{numpy2}
{van der Walt}, S., {Colbert}, S.~C., \& {Varoquaux}, G. 2011, Computing in
  Science and Engineering, 13, 22, \dodoi{10.1109/MCSE.2011.37}

\bibitem[{Virtanen {et~al.}(2020)Virtanen, Gommers, Oliphant, Haberland, Reddy,
  Cournapeau, Burovski, Peterson, Weckesser, Bright, {van der Walt}, Brett,
  Wilson, Millman, Mayorov, Nelson, Jones, Kern, Larson, Carey, Polat, Feng,
  Moore, {VanderPlas}, Laxalde, Perktold, Cimrman, Henriksen, Quintero, Harris,
  Archibald, Ribeiro, Pedregosa, {van Mulbregt}, \& {SciPy 1.0
  Contributors}}]{scipy}
Virtanen, P., Gommers, R., Oliphant, T.~E., {et~al.} 2020, Nature Methods, 17,
  261, \dodoi{10.1038/s41592-019-0686-2}

\bibitem[{{Williams} \& {Berger}(2015)}]{Williams2015ApJ...808..189W}
{Williams}, P.~K.~G., \& {Berger}, E. 2015, \apj, 808, 189,
  \dodoi{10.1088/0004-637X/808/2/189}

\bibitem[{{Williams} {et~al.}(2015{\natexlab{a}}){Williams}, {Berger}, {Irwin},
  {Berta-Thompson}, \& {Charbonneau}}]{Williams2015ApJ...799..192W}
{Williams}, P.~K.~G., {Berger}, E., {Irwin}, J., {Berta-Thompson}, Z.~K., \&
  {Charbonneau}, D. 2015{\natexlab{a}}, \apj, 799, 192,
  \dodoi{10.1088/0004-637X/799/2/192}

\bibitem[{{Williams} {et~al.}(2013){Williams}, {Berger}, \&
  {Zauderer}}]{Williams2013ApJ...767L..30W}
{Williams}, P.~K.~G., {Berger}, E., \& {Zauderer}, B.~A. 2013, \apjl, 767, L30,
  \dodoi{10.1088/2041-8205/767/2/L30}

\bibitem[{{Williams} {et~al.}(2015{\natexlab{b}}){Williams}, {Casewell},
  {Stark}, {Littlefair}, {Helling}, \& {Berger}}]{Williams2015ApJ...815...64W}
{Williams}, P.~K.~G., {Casewell}, S.~L., {Stark}, C.~R., {et~al.}
  2015{\natexlab{b}}, \apj, 815, 64, \dodoi{10.1088/0004-637X/815/1/64}

\bibitem[{{Williams} {et~al.}(2014){Williams}, {Cook}, \&
  {Berger}}]{Williams2014ApJ...785....9W}
{Williams}, P.~K.~G., {Cook}, B.~A., \& {Berger}, E. 2014, \apj, 785, 9,
  \dodoi{10.1088/0004-637X/785/1/9}

\bibitem[{{Zhang} {et~al.}(2020){Zhang}, {Hallinan}, {Brisken}, {Bourke}, \&
  {Golden}}]{Zhang2020ApJ...897...11Z}
{Zhang}, Q., {Hallinan}, G., {Brisken}, W., {Bourke}, S., \& {Golden}, A. 2020,
  \apj, 897, 11, \dodoi{10.3847/1538-4357/ab9177}

\end{thebibliography}
